\documentclass[transmag]{IEEEtran}
\usepackage[utf8]{inputenc}
\usepackage{cite}
\usepackage{url}
\usepackage{color}
\usepackage[normalem]{ulem}
\usepackage{dsfont}
\usepackage[table]{xcolor}

\usepackage{listings}
\usepackage{acronym}
\usepackage{gensymb}
\usepackage{amsmath,amssymb,amsfonts}
\usepackage{algorithmic} 
\usepackage{mathtools}
\usepackage{graphicx} 
\usepackage{mdwmath}
\usepackage{mdwtab}  
\usepackage{textcomp}  
\usepackage[ruled,vlined]{algorithm2e}
\usepackage{enumitem}  
\usepackage[margin=0.75in]{geometry}
\usepackage{wasysym} 
\usepackage{comment}
\usepackage{color,soul}
\DeclareUnicodeCharacter{2061}{}
\DeclareUnicodeCharacter{0301}{}
\DeclareUnicodeCharacter{0015}{}
\usepackage{enumerate}   \DeclareMathOperator*{\argmax}{argmax}
\usepackage{subfigure}

\usepackage{authblk}
\usepackage{array} \newcolumntype{M}[1]{>{\centering\arraybackslash}m{#1}}
\def\BibTeX{{\rm B\kern-.05em{\sc i\kern-.025em b}\kern-.08em
    T\kern-.1667em\lower.7ex\hbox{E}\kern-.125emX}}
\SetKwInput{KwInput}{Input}                
\SetKwInput{KwOutput}{Output} 
\newcommand\norm[1]{\left\lVert#1\right\rVert}
\usepackage{latexsym}

\begin{document}

\title{Low-Complexity Codebook Design for SCMA based Visible Light Communication }

\author{Saumya Chaturvedi,~Dil Nashin Anwar,~Vivek Ashok Bohara,~Anand Srivastava,~Zilong Liu

\thanks{Saumya Chaturvedi, Dil Nashin Anwar, Vivek Ashok Bohara, and Anand Srivastava are with Indraprastha Institute of Information Technology
(IIIT-Delhi), Delhi, New Delhi, 110020, India (e-mail: \{saumyac, dilnashina,
vivek.b, anand\}@iiitd.ac.in). Zilong Liu is with the School of Computer Science and Electrical
Engineering, University of Essex, Colchester CO4 3SQ, U.K. (e-mail:
zilong.liu@essex.ac.uk).}\\}


\IEEEtitleabstractindextext{

\begin{abstract}
Sparse code multiple access (SCMA), as a code-domain non-orthogonal multiple access (NOMA) scheme, has received considerable research attention for  enabling  massive connectivity in future wireless communication systems. 
In this paper, we present a novel codebook (CB) design for SCMA based visible light communication (VLC) system, which suffers from  shot noise.  In particular, we introduce an iterative algorithm for designing and optimizing CB  by considering the impact of shot noise at the VLC receiver.  Based on the proposed CB, we derive and analyze the  theoretical bit error rate (BER) expression for the resultant SCMA-VLC system. The simulation results show that our proposed CBs outperform CBs in the existing literature for different loading factors with much less complexity. Further, the derived analytical BER expression well aligns  with simulated results, especially in  high signal power regions.

\end{abstract}

\begin{IEEEkeywords}
CB Design, Input-Dependent Gaussian Noise (IDGN), Sparse Code Multiple Access (SCMA), Visible Light Communication (VLC).

\end{IEEEkeywords}

}
\maketitle
\section{INTRODUCTION}
In recent years, visible light communication (VLC) has emerged as a reliable alternative to  legacy radio-frequency (RF) based communication, especially for indoor scenarios. VLC offers various advantages such as unlicensed spectrum, free of electromagnetic interference, and ease of deployment using existing illumination infrastructure along with high-level security \cite{singleled_1,singleled_2, single_led3}. Besides, non-orthogonal multiple access (NOMA) has received increasing research attention in the past few years as a key technology for the enabling of massive machine type communication (mMTC) systems, in which a massive number of devices with very high density needs to be served \cite{lisuyu6gpaper_2}, \cite{scma6g_6}. The primary concept behind NOMA is to serve multiple users over the same resource elements (frequency or time slots) using different codebooks (CBs) or power levels. This paper is concerned with the integration of VLC with a disruptive NOMA scheme, called sparse code multiple access (SCMA), where multiple users are served simultaneously with different sparsity structures of the CBs \cite{starqam_5}. Message passing algorithm (MPA) is used for  multi-user detection  by taking  advantage of  the sparse CBs   \cite{starqam_18}. In SCMA, CB design plays a key role in the enhanced error rate performances \cite{starqam_7}. 

To support multiple access in VLC, conventional orthogonal frequency division multiple access (OFDMA) may not be employed as VLC has the requirement of real and positive signal transmission. In order to overcome the above drawback,  direct current-biased optical OFDMA (DCO-OFDMA) and asymmetrically clipped optical OFDMA
(ACO-OFDMA) have been studied. However, DCO-OFDMA suffers from deteriorated  bit error rate (BER), and ACO-OFDMA suffers from reduced spectral efficiency \cite{nomavlc_7}. In comparison, SCMA can be combined with VLC  to increase the spectral efficiency because of the following reasons \cite{nomavlc_importance}:
\begin{itemize}
    
    \item  The dominant line-of-sight (LOS) component and the short distance between light-emitting diode (LED) and photodetector (PD) give rise to a high signal-to-noise ratio (SNR) at VLC receiver, which helps improve the  performance of SCMA.
    
    \item The limited modulation bandwidth of the LEDs can be efficiently utilized by SCMA to achieve high spectral efficiency.
\end{itemize} 



Several research works incorporating SCMA in VLC (SCMA-VLC) systems have been presented in the literature. In \cite{vimal_12}, \cite{vimal_13}, experimental demonstration of SCMA-VLC system with overloading factor of 1.5 was reported \footnote{\color{black}The overloading factor of 1.5 means that the number of users is 1.5 times the number of resource elements.}. 
\color{black}
 In \cite{vimal_14}, a hybrid power-domain SCMA was investigated for VLC, showing improved data rate at the cost of reduced transmission distance. In \cite{vimalpaper}, color-domain SCMA was proposed for VLC for a large throughput gain over conventional white light SCMA system. All the above works used the predefined RF complex SCMA CBs  and made suitable changes (such as Hermitian symmetry and DC-level shifting) so that they can be applied to the VLC system. The authors in \cite{vimal_8} focused on SCMA constellation design for VLC system in the presence of thermal noise with six users and eight resource elements (REs). In \cite{rmedpaper} and \cite{dropaper}, SCMA CB has been designed for three users with four REs considering the impact of varying shot noise in VLC systems. Although \cite{vimal_8,rmedpaper,dropaper} have proposed SCMA CBs for different VLC systems, however,  there is still a lack of literature on designing  SCMA CBs with overloading factor greater than one while considering the presence of shot noise in VLC systems.\\
{Previously, the analytical expression of BER has been discussed} \cite{shotnoise_ber_1,shotnoise_ber_2} {either for power-domain NOMA techniques or for single-carrier scheme (such as on-off keying) for shot noise incorporated VLC systems. However, SCMA is a multi-carrier scheme, and its BER analysis will be different from single-carrier schemes.  Further, when the SCMA-VLC system incorporates varying shot noise, the conventional closed-form BER expression does not follow the simulation results. Therefore, an exact theoretical BER expression for shot noise incorporated SCMA-VLC system has been proposed in this work.
}
\color{black}
To the best of our knowledge, this paper is the first known work that  proposes the SCMA CB design with 1.5 overloading factor for VLC systems considering shot noise.
The  contributions of this paper are summarized as follows:
\begin{itemize}
    \item A novel multidimensional, low-complex, and power-efficient SCMA CB designing technique has been proposed for shot noise incorporated VLC system utilizing the log-sum-exponential operation in the optimization objective function of CB design. The proposed CBs outperform CBs in the existing literature \cite{rmedpaper,dropaper} for
different load factors with much less complexity.
    {\item  The proposed optimization problem and its solution lead to  flexible SCMA CBs which can meet the need of SCMA-VLC systems with varying loading factors.
    }
    \color{black}
    \item The analytical expression of BER for SCMA-VLC system considering the shot noise has been derived. The simulation results show good agreement  with the derived analytical BER expression, especially for higher signal power region.
\end{itemize}

The remainder of the paper is organized as follows. The SCMA-VLC system model is discussed in Section II. The proposed CB design and the theoretical BER expression for SCMA over shot noise VLC system are discussed in Section III. The results are presented and analyzed in Section IV. Section V concludes the paper.

\underline{\emph{Notations:}} Throughout this paper, x, \textbf{x}, \textbf{X} denote a scalar, vector and a matrix, respectively. Symbols $\textbf{x}^{T}$  and $\textbf{X}^{T}$ represent transpose of \textbf{x} and \textbf{X}, respectively. The ${i}$th element of vector $\textbf{x}$ is denoted by ${x}_i$ and $(\textbf{X})_{ij}$ denotes the $i$th row and $j$th column element of matrix \textbf{X}. Symbols $\mathbb{B}$  and $\mathbb{R}_{>0}$ 
represents the set of binary numbers and, real and positive numbers, 
respectively. The diagonal matrix is denoted by $\text{diag(\textbf{x})}$ where the $i$th diagonal element is ${x}_i$  
and `log(\textbf{x})' denotes the natural logarithm of each element of $\textbf{x}$, respectively. The maximum value of $f(x)$ as $x$ is varied over all its possible values is denoted by `$\max \limits_{x} f(x)$'. The probability density function (PDF) of a Gaussian random variable (RV) with mean $\mu$ and variance $\sigma^2$ is denoted by  $\mathcal{N} [\mu,\sigma^2]$. The trace of a square matrix $\textbf{A}$ is denoted by $\text{Tr}(\textbf{A})$  which is calculated by the sum of main diagonal elements of $\textbf{A}$. The selection of $r$ elements from a set of $n$ elements is denoted by 
$\binom{n}{r}$
and $\norm{\textbf{x}}$ denotes the L2 norm of $\textbf{x}$, respectively.

\section{System Model}
\subsection{Input-dependent Gaussian noise}
\begin{figure*}[htbp]
\centering
\includegraphics[scale=0.4,trim=0.1cm 1cm 0cm 0.5cm,clip]{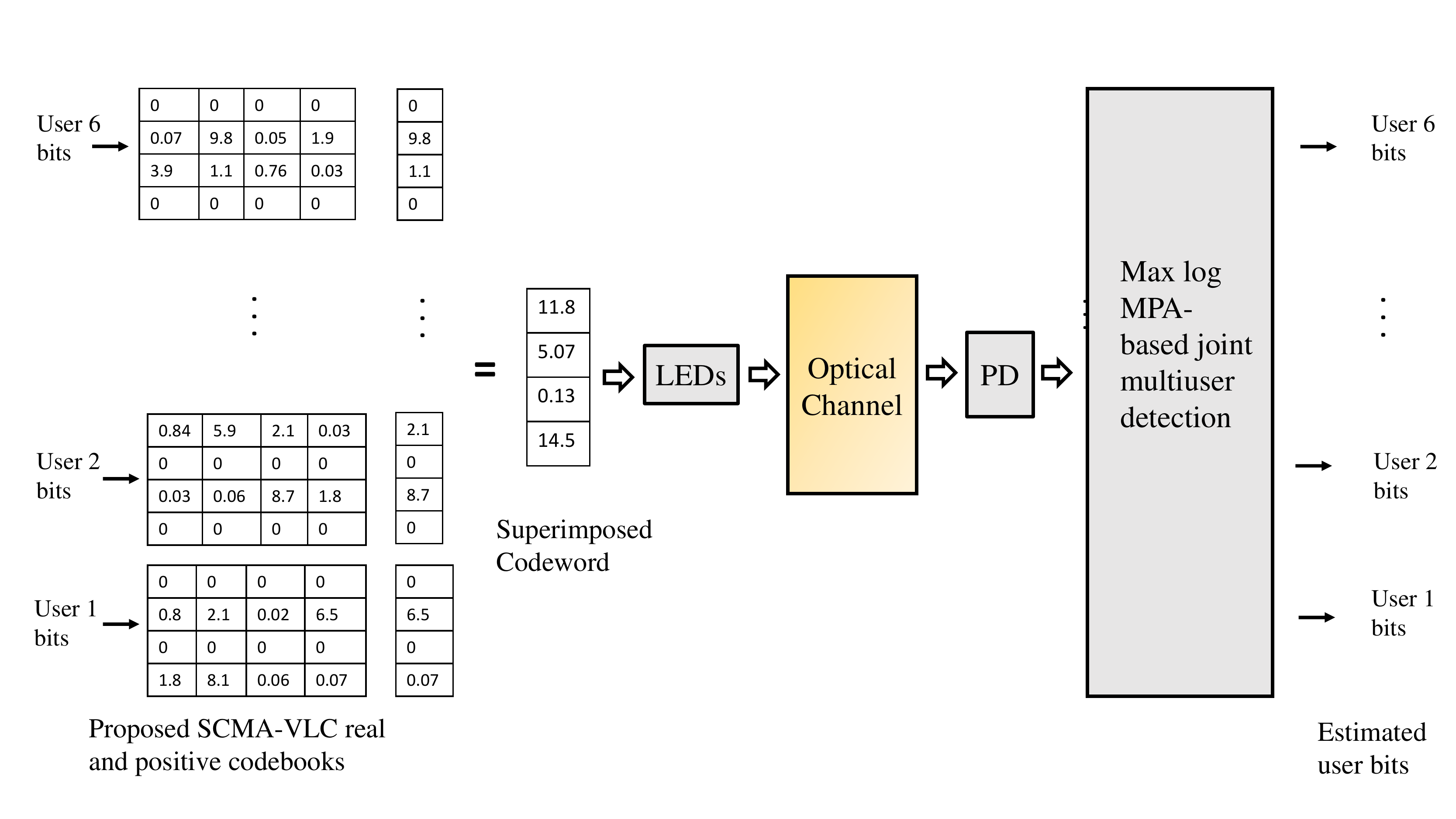}\\
\caption{ The system model diagram of $4\times 6$ SCMA with the proposed CBs at the transmitter in VLC channel.}
\label{fig:scmasystemmodel}
\end{figure*}

In the VLC system, the optical receiver (i.e., PD) measures the intensity of the incident optical signal and converts it into electrical signal.
The interaction of the incoming photons with the matter of PD is a statistical process \cite{dna_rv2}. The fluctuations in the number of photons detected, results in fluctuating current which cause shot noise at the optical receiver. The generated shot noise  depends on the incident optical signal itself \cite{sdnscalingfactorpaper,yaseenidgn11_23,yaseenidgn11}. Further, the thermal motion of the electronic carriers result in fluctuating voltage known as thermal noise.
\color{black}
{ In case of strong ambient light, the impact of thermal noise is more and the shot noise becomes very low. Thus, the 
noises at the VLC system are assumed to be independent of the received signal} \cite{shotnoise2_4,shotnoise2_5,shotnoise2_9}.{ However, in case
of indoor scenario, the distance between transmitter and receiver is less and the dominant signal is LOS} \cite{dna_los_rv2}, {thus the received power  is large. However, in practical VLC systems, typical illumination and communication scenarios offer very high SNR}\cite{shotnoise2_11,shotnoise2_12}. {In case of indoor scenario (as considered in our work), since the received power is large, the shot noise is more dominant and cannot be neglected.} 
\color{black}
The variance of shot noise is given as
\begin{align}
    \sigma_{sh}^2=2 q x_p \mathfrak{R} \text{B} + 2q I_B I_2 \text{B},
\end{align}
where $q$ is the electronic charge, $x_p$ is the intensity of incident optical signal, $\mathfrak{R}$ is responsivity, $I_B$ is the background noise current,
and $I_2$ is the noise-bandwidth factor, and $\text{B}$ is the effective bandwidth of the optical receiver.  \\
The thermal noise is assumed to be Additive white Gaussian noise (AWGN), which is independent of the intensity of the incident signal \cite{sdnscalingfactorpaper}.   The shot noise affected optical intensity channel can be distinguished as an input-dependent Gaussian noise (IDGN), where variance of IDGN can be used to characterise the physical properties of the VLC system \cite{sdnscalingfactorpaper, vlc_idgn2}. Consequently, the received signal $y$ can be expressed as
\begin{align}
    y=x_p+\sqrt{x_p}Z_1 +Z_0,   ~~~~~~~y \in \mathbb{R}, ~x_p\geq 0,
\end{align}
where $Z_1 \sim \mathcal{N}_\mathbb{R} [0,\varsigma^2\sigma^2]$ denotes a real Gaussian RV with zero mean and $\varsigma^2\sigma^2$ variance describing IDGN (shot noise), and $Z_0 \sim ~\mathcal{N}_\mathbb{R} [0,\sigma^2]$ describes the thermal noise. Here, $Z_1$ and $Z_0$ are assumed to be independent, and $\varsigma^2 \in [1,10]$ denotes the strength of shot noise  with respect to thermal noise. 
\color{black}
{In an indoor scenario,  as the transmitter and the receiver parameters change, the received power changes, and thus the shot noise varies. In this paper, the variation in shot noise has been depicted by considering different shot noise factor
$\varsigma^2$ values.}
\color{black}
\subsection{The SCMA-VLC Model}
\label{sec_2b}
Consider a downlink 
\color{black}{indoor} 
\color{black}
LOS VLC system where multiple users are being served, each equipped with a single receiving device. Every $J$ users are grouped together for SCMA, and each group is supported by $K$ REs (e.g., time or frequency slots), providing the load factor $\lambda=J/K$. In Fig.\ref{fig:scmasystemmodel}, the data transmission and reception of six users ($J=6$) on four REs ($K=4$) with $\lambda=1.5$ is shown.

At the transmitter, $b=\text{log}_2 M$ bits are mapped to $K$-dimensional real and positive codewords, with  $M$ number of codewords in a CB. 
These codewords are sparse vectors with $N$ non-zero elements, $K>N$. Each user has a dedicated CB allotted to it. The CB of $j$th user can be expressed as
\begin{align}
    \mathbb{A}_j={\textbf{V}_j \textbf{C}_j }, ~~~~~~~~~~~~~\text{for}~ j=1,\cdots,J,
\end{align}
where $\textbf{V}_j \in \mathbb{B}^{K \times N}$ and $\textbf{C}_j \in \mathbb{R}_{>0}^{N \times M}$ denote the mapping matrix and constellation matrix of the $j$th user, respectively. The constellation matrix $\textbf{C}_j=[\textbf{c}_j^1,\cdots,\textbf{c}_j^M]$, where $\textbf{c}_j^m \in \mathbb{R}_{>0}^{N \times 1}$ denotes the $m$th constellation point of $\textbf{C}_j$.  The whole CB structure of SCMA can be represented by the factor graph matrix, $\textbf{F}=(\textbf{f}_1,\textbf{f}_2,\cdots,\textbf{f}_J)$, where $\textbf{f}_j=\text{diag}(\textbf{V}_j \textbf{V}_j^T)$. The user $j$ is being served by resource node (RN) $k$, if $\textbf{F}_{kj}=1$ $(1 \leq k \leq K, 1 \leq j \leq J)$. For a $4 \times 6$ SCMA block having $K=4$ REs and $J=6$ users, one example of the factor graph  matrix of size $4 \times 6$ is given as  
\begin{align}\label{fac_gra_4_6}
    \textbf{F}_{4 \times 6}=
    \begin{bmatrix}
    0&1&1&0&1&0\\
    1&0&1&0&0&1\\
    0&1&0&1&0&1\\
    1&0&0&1&1&0\\
    \end{bmatrix}.
\end{align}

\begin{figure}[htbp]
\centering
\includegraphics[scale=0.3,trim=3.25cm 4.75cm 4.5cm 6.5cm,clip]{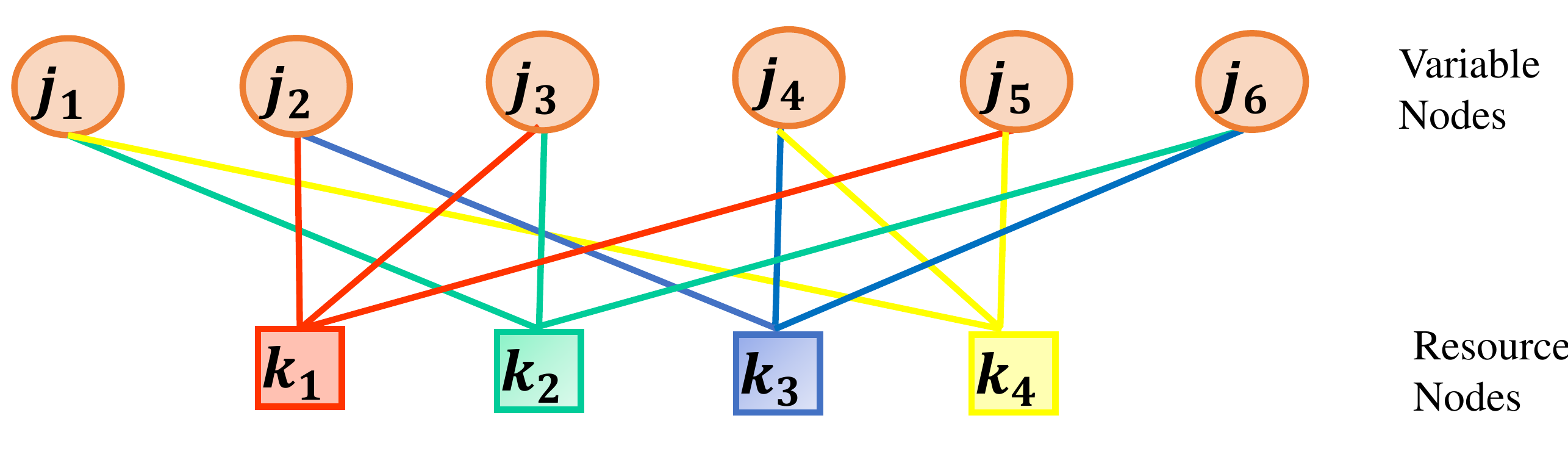}
\caption{Factor graph representation of $4 \times 6$ SCMA system with $d_f = 3$ and $N = 2$.}
\label{fig:fact_gr}
\end{figure}

Fig. \ref{fig:fact_gr} shows an example of factor graph representation of six users (or Variable Nodes (VNs)) with four RNs, $N$=2 and data of $d_f$ users is being superimposed on each RN ($d_f=3$ for $\textbf{F}_{4 \times 6}$ shown in (\ref{fac_gra_4_6})).  The binary mapping matrices corresponding to the six users are given below
\begin{equation}\label{mappingmatrix}
\begin{split}
    \textbf{V}_1 =
\begin{bmatrix}
0 & 0\\
1 & 0\\
0 & 0\\
0&1\\
\end{bmatrix}, 
\textbf{V}_2 =
\begin{bmatrix}
1 & 0\\
0 & 0\\
0 & 1\\
0 & 0\\
\end{bmatrix}, 
\textbf{V}_3 =
\begin{bmatrix}
1 & 0\\
0&1\\
0 & 0\\
0 &0\\
\end{bmatrix}, 
\\\\
\textbf{V}_4 =
\begin{bmatrix}
0&0\\
0 & 0\\
1 &0\\
0&1\\
\end{bmatrix},
 \textbf{V}_5 =
\begin{bmatrix}
1 & 0\\
0 & 0\\
0 & 0\\
0 & 1\\
\end{bmatrix},
\textbf{V}_6 =
\begin{bmatrix}
 0 & 0\\
1 & 0\\
0&1\\
0 &0\\
\end{bmatrix}.
\end{split}
\end{equation}
The set of CBs for $J$ users is dependent on the constellation matrices $\mathbf{C}_j, ~j=1,\cdots,J $ as the mapping matrices remain fixed as shown in (\ref{mappingmatrix}). Assuming synchronous multiplexing between users, the received signal can be expressed as
\begin{equation}\label{rec_sig_eqn}
\begin{split}
&\textbf{y} = \sum_{j=1}^{J} \text{diag} (\textbf{h}_{j}) \textbf{x}_{j} + \text{diag}\left(\sum_{j=1}^{J} \text{diag} (\textbf{h}_{j}) \textbf{x}_{j}\right)^{\frac{1}{2}}\textbf{n}_{sh}+\textbf{n}_{th},\\
&= \sum_{j=1}^{J} \text{diag} (\textbf{h}_{j}) \textbf{V}_{j} \textbf{c}_j + \text{diag}\left(\sum_{j=1}^{J} \text{diag} (\textbf{h}_{j}) \textbf{V}_{j}\textbf{c}_j\right)^{\frac{1}{2}}\textbf{n}_{sh}+\\& \textbf{n}_{th},
\end{split}
\end{equation}
where $\textbf{x}_j \in \mathbb{R}_{>0}^{K \times 1}$ is a codeword of $j$th user, $\text{diag}(\textbf{h}_j)$ denotes the channel matrix of $j$th user, the second term represents the IDGN with $\textbf{n}_{sh} \sim \mathcal{N}_\mathbb{R} [\textbf{0},\varsigma^2\sigma^2\textbf{I}]$, and $\textbf{n}_{th} \sim \mathcal{N}_\mathbb{R} [\textbf{0},\sigma^2\textbf{I}]$. The channels are assumed to be unit vectors for simplicity, and the extension to the general VLC values is straight-forward.
\color{black}
{In case of the movement of the users, the channel characteristics of the users $((\textbf{h}_j )$ of $j$th user in $(\ref{rec_sig_eqn}))$ will be affected. The random
waypoint (RWP) mobility model is a simple and straightforward
stochastic model that describes a human 
movement behavior in indoor scenarios} \cite{anandusermobility_15}. In case of user mobility, the received signal expression remains same as (\ref{rec_sig_eqn}), {however the channel vector ($\textbf{h}$) varies as per the RWP model in} (\ref{rec_sig_eqn}). 
 
 \color{black}
\subsection{SCMA-VLC  Decoding}
Assuming perfect channel estimation and the set of CBs known to receiver, the SCMA multi-user codeword $\textbf{X}=[\textbf{x}_1,\textbf{x}_2,\cdots,\textbf{x}_J]$ can be detected by solving the joint  maximum a-posteriori (MAP)  probability mass function (PMF) of the transmitted codewords \cite{starqam_18},  \cite{superimposedcwmed2021_30}. The estimated $\textbf{X}$ can be written as 
\begin{align} \label{multiuser_codeword}
    \hat{\textbf{X}} = \argmax_{\textbf{x}_j \in \mathbb{A}_j,\forall j} p(\textbf{X} \vert \textbf{y})\,,
\end{align}
where $ \mathbb{A}_j$ represents the set of codewords allotted to the $j$th user. The marginal distribution of (\ref{multiuser_codeword}) with respect to $\textbf{x}_j$ (i.e., summing the joint PMF of multi-user codeword over all values of $\textbf{X}$ except $\textbf{x}_j$) leads to $j$th user detected symbol as

\begin{align}
    \hat{\textbf{x}}_j = \argmax_{\textbf{x}_j \in \mathbb{A}_j} \sum_{{\sim} \textbf{x}_j} p(\textbf{X} \vert \textbf{y})\,.
\end{align}
Using Bayes’ rule
\begin{align}\label{map1}
    \hat{\textbf{x}}_j = \argmax_{\textbf{x}_j \in \mathbb{A}_j} \sum_{{\sim} \textbf{x}_j} f(\textbf{y} \vert \textbf{X})  P(\textbf{X})\,,
\end{align}
where $P(\textbf{X})$ denotes the joint a-priori PMF of all transmitted codewords and $f(\textbf{y} \vert \textbf{X})$ denotes the conditional pdf of the received vector. 
Assuming noise components  are independent and identically distributed for all REs. Then (\ref{map1}) becomes
\begin{align}\label{mpf}\nonumber
    \hat{\textbf{x}}_j = \argmax_{\textbf{x}_j \in \mathbb{A}_j} \sum_{{\sim}\textbf{x}_j} \left(P(\textbf{X}) \prod_{k=1}^{K} f(y_k \vert \textbf{X})\right)\,, 
    \enspace \\\text{for}~ j=1, \cdots, J,
\end{align}
where $y_k$ denotes the received signal at the $k$th RE. Solving the above marginal product of function (MPF) problem with brute force will lead to exponential complexity. Taking  advantage of sparse CBs, this problem can be approximately solved by iterative MPA with moderate complexity \cite{starqam_18}, \cite{mpacomplexity}. MPA works as a 
decoding algorithm based on passing the extrinsic information between RNs and VNs. The first message is passed from the leaf node (node which has only one neighboring node), and a node of degree $d$ will remain idle until it has received messages from the $d-1$ number of neighboring nodes. MPA involves a large number of exponential operations which are of high complexity. A mathematical simplification is to move MPA to logarithm domain and use Jacobian logarithm, which gives a simpler version called Max-Log-MPA \cite{scmapotentialchallenges_109}. With Jacobian logarithm, the log-sum of exponentials can be approximated as a maximum operation as shown below:
\begin{align}\label{jacob}
    \log (\exp (a_1)+\cdots+\exp(a_J)) \approx \text{max}(a_1,\cdots,a_J).
\end{align}
The Max-Log-MPA for detection of SCMA codewords in VLC is based on three steps as shown in \textbf{Algorithm 1}. Let $\zeta_j$ and $\xi_k$, be the set of nodes connected to VN $j$ and RN $k$, respectively. Let $\eta_{k \rightarrow j}$ and $\eta_{j \rightarrow k}$ be the message passed from RN $k$ to VN $j$ and vice-versa. Also, $r \in \xi_k \setminus \{j\}$ denotes all the VNs of $\xi_k$ except VN $j$ and $d \in \zeta_j \setminus \{k\}$ denotes the RNs in $\zeta_j$ except the $k$th RN.
In \textbf{Algorithm \ref{maxlogmpa}}, 
it is assumed that the received signal \textbf{y} and all the CBs  \textbf{C}$_j$, $\forall j=1, \cdots,J$ are known at the receiver. The total number of iterations considered in $\textbf{Step 2}$ are $N_{t}$ and, VNs and RNs are updated for each iteration $t$. Also, the channel coefficient between $k$th RN and $m$th VN is denoted by $h_{km}$ and the codeword element transmitted by $m$th VN on $k$th RN is denoted by $x_{km}$, respectively.

\begin{algorithm}
  \caption{Max-Log-MPA}\label{maxlogmpa}
  \footnotesize
  \KwInput{\textbf{y}, \textbf{C}$_j$, $\forall j=1, \cdots,J$.}
  \KwOutput{Estimated value of bits for $j$th user, $\forall j=1,\cdots,J.$}
  \begin{algorithmic}[1]
    \raggedright
  \STATE{\textbf{Step 1: Initialization}\\
  The prior probability of each codeword: $\eta_{j \rightarrow k}^0=\text{log}\frac{1}{M}.$
  }
  \STATE {\textbf{Step 2: Message passing along the nodes}} \\
  
  \For{$t =1~\KwTo~N_{t}$}
{a) Update RN:
\begin{align*}
    &\eta_{k \rightarrow j}^t(\textbf{x}_j) =\\& \max_{{\sim}\textbf{x}_j} \Bigg( \frac{-1}{2\rho^2} \norm{y_k- \sum_{m \in \xi_k} h_{km}x_{km}}^2 +\sum_{r \in \xi_k \setminus \{j\}} \eta_{r \rightarrow k}^{t-1}(\textbf{x}_r)\Bigg),\\
    \end{align*}
    where, $\rho^2= \sigma^2+\varsigma^2 \sigma^2 \sum_{m \in \xi_k} h_{km}x_{km}.$\\
    b) Update VN:
\begin{align*}
    \eta_{j \rightarrow k}(\textbf{x}_j) =\text{log}\left(\frac{1}{M}\right)+\sum_{d \in \zeta_j \setminus \{k\}} \eta_{d \rightarrow j}^{t-1}(\textbf{x}_j).
\end{align*} \\
\textbf{end} } 

 \STATE {\textbf{Step 3: Bit log-likelihood ratio (LLR)}\\
  The final belief computed at each VN is:}
  \begin{align*}
     \text{log}~ \textbf{I}_j(\textbf{x}_j)= \text{log}\left(\frac{1}{M}\right)+\sum_{k \in \zeta_j } \eta_{k \rightarrow j}(\textbf{x}_j)
  \end{align*}
{{The $\kappa$th bit LLR at $j$th VN is }
\begin{align*} \label{bit_llr}
 &LLR_{\kappa}^j =\\& \max_{\{\textbf{x}_j \in \mathbb{A}_j |\kappa=0\}}     (\text{log}{( \textbf{I}_j(\textbf{x}_j)}))-\max_{\{\textbf{x}_j \in \mathbb{A}_j |\kappa=1\}}     (\text{log}{( \textbf{I}_j(\textbf{x}_j)})).
\end{align*}
  where, $\hat{\kappa}_j=0$ if $LLR_{\kappa}^j$ is positive otherwise $\hat{\kappa}_j=1$.
  }
  \end{algorithmic}
\end{algorithm}
In the case of shot noise, the variance at each RN is different and dependent on the superimposed codeword element at the corresponding RN, while in the presence of only thermal noise,  $\varsigma^2=0$, so $\rho^2=\sigma^2$ and variance will be the same for all RNs.
In case of three users scenario, the factor graph would not be cyclic and number of elements being superimposed on each RN would be different ($d_f=2$ for $k=1,2$ and $d_f=1$ for $k=3,4$). Since $d_f$ is different for different REs, therefore,  \textbf{Algorithm 1} will change accordingly \cite{factorgra_sumprodalgo}.

The computational complexity of standard MPA and Max-Log-MPA considering the effect of shot noise in decoding is shown in Table \ref{table:1} \cite{singleled_23}.
The number of operations is shown for the update step of RN as shown in \textbf{Step 2} (a) of \textbf{Algorithm 1}. Max-Log-MPA requires  more addition operations and  less multiplication operations than MPA. This gap of increase in addition and decrease in multiplication operations between MPA and Max-Log-MPA increases with the overloading factor.

\begin{table}[htbp]
\caption{Complexity analysis of MPA and Max-Log-MPA considering shot noise.}
\centering
\begin{tabular}{|c|c|c|} 
\hline 
\textbf{Operations}& \textbf{MPA} & \textbf{Max-Log-MPA} \\ \hline\hline
Exponential & $M^{d_f}K d_f N_{it} $ & 0 \\ \hline
Multiplication & $(d_f+3)M^{d_f}K d_f N_{it}$ & $4 M^{d_f}K{d_f}N_{it}$  \\ \hline
Addition & $(2{d_f}+2)M^{d_f}K{d_f}N_{it} $ &  $(3d_{f}+1)M^{d_f}K{{d_f}^2}N_{it}$ \\ \hline
Comparison & 0 & $ M^{d_f}K{{d_f}}N_{it}$ \\ \hline

\end{tabular}
\label{table:1}
\end{table}

\section{CB Design and Optimization}
For SCMA CB design and optimization, most researchers aim for maximization of the minimum Euclidean distance (MED) between the superimposed codewords \cite{starqam,klimenyvmed_genetic,cbsurvey_papr}.  
In this section, a method of designing constellation matrices $\textbf{C}=[\textbf{C}_1,\cdots,\textbf{C}_J] \in \mathbb{R}_{>0}^{N \times MJ}  $ is introduced and discussed as an optimization problem.

\subsection{The Problem Formulation}
In the AWGN channel, MED is taken as one of the key performance indicators for determining the performance of the designed SCMA CB. Let $\textbf{s}_i$ be a superimposed codeword from the combined constellation matrix 
$\mathbb{M}_v \in \mathbb{R}_{>0}^{K \times M^J}$. Then, the MED between $\textbf{s}_i \in \mathbb{M}_v$ and $\textbf{s}_j \in \mathbb{M}_v$ is given as
\begin{align}
    d'_{ij}=(\textbf{s}_i-\textbf{s}_j)^T (\textbf{s}_i-\textbf{s}_j), ~~~~~~~~~ \forall~ i<j \in [1,M^J].
\end{align}
The MED emerge as a critical parameter because of the pair-wise error probability (PEP) for AWGN channel which is given as
\begin{align}\label{pepawgn}
        P'(\textbf{s}_i \rightarrow \textbf{s}_j)=Q \left(\sqrt{\frac{\norm{\textbf{s}_i-\textbf{s}_j}^2}{2\sigma^2}}\right).
\end{align}
However, in case of IDGN, the PDF of the received signal $\textbf{y}$ is  given as (assuming unit channel gain)
\begin{align}\label{pepidgn}
f(\textbf{y}\vert \textbf{s}_i)= \frac{1}{(2\pi)^{K/2}} \frac{1}{\vert{\Sigma_{i}}\vert^{\frac{1}{2}}} \text{exp}\{-\frac{1}{2}(\textbf{y}-\textbf{s}_i)^T \Sigma_i^{-1} (\textbf{y}-\textbf{s}_i)\}
,
\end{align}
where $\Sigma_i=\text{diag}\{\varsigma^2\sigma^2 \textbf{s}_i+ \sigma^2 \textbf{1} \}$ denotes the covariance matrix for the $\textbf{s}_i$th superimposed codeword.  It can be seen from (\ref{pepidgn}) that the variance is now dependent on the superimposed codeword. 

In the case of only thermal noise, MED is taken as a critical parameter for designing CB where variance is independent of the superimposed codeword. However, in the case of shot noise, when variance is no longer independent, the conventional MED cannot serve as an optimal performance indicator.  With the effect of shot noise, the shape of constellation points gets distorted.
 To mitigate the effect of IDGN, the authors in \cite{rmedpaper} introduced an improved optimization objective based on Rotated-MED  for designing constellation matrices. The rotated Euclidean distance (RED) between superimposed codewords $\textbf{s}_i$ and $ \textbf{s}_j$ can be expressed as 
\begin{multline}\label{red}
        d_{ij}=(\textbf{s}_i-\textbf{s}_j)^T  {\textbf{G}_i}^{-\frac{1}{2}} {\textbf{G}_j}^{-\frac{1}{2}}(\textbf{s}_i-\textbf{s}_j),\\ ~~ \forall~ i<j \in [1,M^J],
\end{multline}
where $\textbf{G}_i=\varsigma^2 \text{diag} \{\textbf{s}_i\}+\textbf{I}$, $\textbf{G}_j=\varsigma^2 \text{diag} \{\textbf{s}_j\}+\textbf{I}$ and  $\textbf{I} \in \mathbb{B}^{K \times K} $ represents the Identity matrix. It is to be noted from (\ref{red}) that the new parameter RED incorporates the shot noise effect while computing the Euclidean distance between superimposed codewords. The authors in \cite{rmedpaper} aimed at maximizing the rotated minimum Euclidean distance (R-MED) subject to the power constraints. The objective function proposed in \cite{rmedpaper} for SCMA CB design is given as
\begin{align}\label{maximin}
    \max_{\textbf{L} \in \mathbb{R}_{>0}^{NMJ \times 1} } \left\{ \min_{1\leq i \leq j \leq M^J} d_{ij}(\textbf{L})\right\},
\end{align}
where $\textbf{L}=\text{vec}(\textbf{C})$. This was further simplified as 
\begin{align}\label{min_rmed_objfn}
    \max_{\textbf{L} \in \mathbb{R}_{>0}^{NMJ \times 1} ,~ d_\text{min $>$ 0}}  d_\text{min},
\end{align}
where $d_\text{min}$ denotes the minimum value of RED. It has  been verified in \cite{rmedpaper} that R-MED based CB outperforms the MED based CB for VLC systems affected by shot noise. The results in \cite{rmedpaper} have shown that  for $J=3, K=4, M=4, \varsigma^2=5$ and  $\sigma^2=0.01 $, BER of $10^{-3}$ is achieved at $P_e =10$ and $P_e=7$ for MED and R-MED based CBs, where $P_e$ denotes the maximum  electrical power allotted to a user. Since from the literature we can understand that RED works as an important parameter for designing CB than MED, we have included RED in our optimization problem. Further, the authors in \cite{dropaper} introduced the concept of distance range (DR) based on the maximum and the minimum values of the RED. They have mentioned that for the low to medium SNR region, the BER is no longer dependent on just R-MED. The optimization problem proposed in \cite{dropaper} for CB designing is 
\begin{subequations}

\label{maximindro}

\begin{align}
\max_{ d_\text{min}>0, d_\text{max}>0 } \left\{  d_\text{min}-\alpha d_\text{max}\right\} 
\label{subeq1}& \\
~~~~\textrm{s.t.} ~~~~ L_i >0, ~~~~~~~~~~~~~ \forall ~ i \in [1,NMJ], \label{subeq2} & \\
   \frac{1}{M} \text{Tr} ( \textbf{C}_j^T \textbf{C}_j) \leq P_e, ~~~~\forall ~j \in [1,J],   \label{subeq3} & \\
   d_{ij} \geq d_{\text{min}},  ~~~~\forall ~ i<j, \label{subeq4} &\\
   d_{ij} \leq d_{\text{max}},  ~~~~\forall ~ i<j, \label{subeq5} &
\end{align}
\end{subequations}
where $d_\text{max}$ denotes the maximum value of RED, and $\alpha >0$ denotes the weighting factor. The objective function  shown in (\ref{subeq1})  is dependent on  $d_\text{min}$ and $d_\text{max}$. The real and positive optical signal constraint is considered in (\ref{subeq2}) and maximum electrical power constraint is considered in (\ref{subeq3}). All the REDs will lie between  $d_\text{min}$ and $d_\text{max}$ (\ref{subeq4}, \ref{subeq5}). Also, it is to be noted that, lesser the overlapping between the constellation points of a user, better the performance of the designed CBs.

\subsection{The Objective Function}
The objective function mentioned in (\ref{maximin}) considers all the REDs while designing and optimizing the CBs to get optimum results. However, considering all the REDs makes the objective function non-convex and non-differentiable \cite{dnaaccess_7}. To solve (\ref{maximin}), authors in [17] have proposed the objective function as shown in (\ref{maximindro}). 
It should be noted that the numerical methods used for solving
the nonlinear maxi-min problem aim at approximating the
$ F(x)= \min_i f_i(x)$ by close enough smooth functions, and the closeness of approximation decide the nearness to optimal solutions \cite{dnaaccess_22}. If we consider $\sum_i \phi(f_i(x))$,  the approximation to $F(x)$ becomes better when $\phi(.)$ is a strong convex function (i.e., large positive second derivative).
One approach to solve this non-convex problem is to use well known ``log sum-exponentials" continuous approximation \cite{dnaaccess_16}, \cite{dnaaccess_11}. With this approximation, the objective function in (\ref{maximin}) transforms to
\begin{align}\label{logsum1}
    \min_{\textbf{L} \in \mathbb{R}_{>0}^{NMJ \times 1} } \frac{1}{\beta} \text{ln} \left\{\sum_{i \neq j} e^{-\beta~ d_{ij}(\textbf{L})} \right\},
\end{align}
where $\beta >0$ is a tuning parameter which results in the smoothness of  (\ref{logsum1}) and affects the accuracy of the model. 

\begin{itemize}
    \item 
\textcolor{black} {Example 1: Let $\beta=1$, and there are two distances $d_1 \gg d_2$, then\\
$  - \text{ln}(e^{-d_1} + e^{-d_2}) \\  \approx -\text{ln}(e^{-d_2}) =d_2 = \min \{d_1,d_2\}.$}

\end{itemize}
The function in (\ref{logsum1}) is differentiable but still non-convex; however, optimization algorithms based on the derivation of the Hessian of the objective function can find the local optimal solution \cite{dnaaccess_14}. Also, if the objective function is optimized number of times with initialization from small set of random numbers and choosing the $\beta$ parameter wisely, the likelihood of obtaining the global optimal solution increases \cite{dnaaccess}. Thus, the optimization problem is formulated as

\begin{subequations}\label{maximinlogsumexp}
\begin{align}
\min_{\textbf{L} \in \mathbb{R}_{>0}^{NMJ \times 1} } \quad  \frac{1}{\beta} \text{ln} \left\{\sum_{i \neq j} e^{-\beta~ d_{ij}(\textbf{L})} \right\}  \label{subeq11} \\
~~~~\textrm{s.t.} ~~~~ L_i >0, ~~~~~~~~~~~~~ \forall ~ i \in [1,NMJ], \label{subeq12} & \\
   \frac{1}{M} \text{Tr} ( \textbf{C}_j^T \textbf{C}_j) \leq P_e, ~~~~\forall ~j \in [1,J],   \label{subeq13} 
\end{align}
\end{subequations}
where (\ref{subeq12}) denotes the real positive intensity constraint of the transmitted codeword elements and (\ref{subeq13}) denotes that the average electrical power of the generated codewords  should be less than the maximum electrical power ($P_e$) allotted per user. The optimization problem in (\ref{maximinlogsumexp}) can be solved using commercially available solvers  like MATLAB's $fmincon$. The procedure for solving  the above optimization problem is summarized in \textbf{Algorithm 2}.

\begin{algorithm}
\caption{SCMA CB Optimization Algorithm}     
\label{alg:1}
\footnotesize
\textbf{Step 1: Initialization}:\\{ Randomly initialize $\textbf{L} = \textbf{L}^0$ satisfying the intensity and power constraints, $\beta=1$.}\\
\textbf{ Step 2: Optimization}
\label{algo:on-grid2}
\begin{algorithmic}[1]
\raggedright
 \FOR{$\beta \leq 30$}

 \STATE{(a) Solve the optimization problem
$[f_v^{t'},\mathbf{L}^{t'}] \gets \min \mkern15mu \frac{1}{\beta} \text{ln} \left\{\sum_{i \neq j} e^{-\beta~ d_{ij}(\textbf{L})} \right\}$ \par
\vspace{0.5em} $\mkern100mu$ \text{subject to Constraints.}\\
where $f_v^{t'}$ denotes the value of the objective function at the obtained solution $\textbf{L}$ at $t'$th iteration. }
\STATE{(b) \IF{ $ \vert{{f_v}^{t'}- {f_v}^{{t'}-1}} \vert \geq 10^{-3};$} 
 \STATE{\emph{t'=t'}+1,} go to \textbf{Step 2} (a);
  \ELSE
\STATE{exit}
\ENDIF}
 \ENDFOR
 
 \KwOutput{$\mathbf{L}^{t'}$ as the SCMA output constellation stacked vector.}
 
\end{algorithmic}
\end{algorithm}

\subsection{Complexity Analysis}
For the case of the proposed optimization problem in (\ref{maximinlogsumexp}), initially, the set of all multiplexed vectors (or superimposed codewords) $\mathbb{M}_v$ 
is computed. With $J$ users and $M$ number of codewords in the CB, the number of superimposed codewords obtained are $M^J$, i.e., $\vert \mathbb{M}_v \vert = M^J$. Since the complexity of operations such as addition and subtraction is $O(1)$ therefore, the complexity of computing multiplexed vectors is $O( J M^J)$. Next, for generating the covariance matrix ($\Sigma_i$ for $\textbf{s}_i$ superimposed codeword), each multiplexed vector is multiplied with the scalar $\varsigma^2$ which results in the complexity $O(K M^J)$. The complexity of computing RED (as shown in (\ref{red})) between two multiplexed vectors  is $O(K)$ as the length of each vector is $K$ and the total number of multiplexed vectors are $M^J$. Thus, the total number of REDs between $M^J$ multiplexed vectors is  
$\binom{M^J}{2}$
and the total complexity for calculating REDs will be $O(K~ {\binom{M^J}{2}})$. Next, the complexity of computing the objective function (\ref{subeq11}) is 
$\binom{M^J}{2}$. 
The complexity of constraints in (\ref{subeq12}) and (\ref{subeq13}) is $O(1)$ since it includes addition and  square operations. The maximum value of $\beta$ at which the algorithm converges is denoted by $\beta_{max}$ and number of iterations for the MATLAB's $fmincon$ solver is $N_{it}$, then total complexity to generate the constellation matrices for $J$ users using \textbf{Algorithm 2} is $O_{Logsum}=[\beta_{max} N_{it} (J M^J+ K M^J+ {K~ {\binom{M^J}{2}}} + {\binom{M^J}{2}} + 1 ) ]$. If we take the most dominant part resulting for complexity, it is equal to $O(\beta_{max} N_{it} K~ {\binom{M^J}{2}} )$. 

In case of the DR based CB \cite{dropaper} optimization problem (as shown in (\ref{maximindro})), the total complexity comes from the computation of the objective function and four constraints. The third and fourth distance based constraints (\ref{subeq4}, \ref{subeq5}) are non-convex and are approximated using first order Taylor series \cite{rmedpaper}. From the calculation of multiplexed vectors, covariance matrices and REDs, the complexity is $O(J M^J+K M^J+{K~ {\binom{M^J}{2}}})$. The complexity of intensity and power constraints (\ref{subeq2}, \ref{subeq3}) is $O(1)$. For the approximation of two non-convex constraints (\ref{subeq4}, \ref{subeq5}), firstly the coefficient matrix $\Pi_i$ is calculated to obtain multiplexed vector $\textbf{s}_i$ from $\textbf{L}$ \cite{rmedpaper}, and its complexity is $M^J KNMJ$. Next, we found that the derivative of RED has the highest complexity because of the kronecker product operation involved. The derivative term has the complexity $O( { (KNMJ)^2 ~ {\binom{M^J}{2}}})$. So, with $N_{out}$ as total number of outer loop iterations, the total complexity is $ O_{DR}= [O(N_{out} N_{it} (J M^J+ K M^J+ {K~ {\binom{M^J}{2}}} +  {\binom{M^J}{2}} + 1 +M^J KNMJ+{ (KNMJ)^2 ~ {\binom{M^J}{2}}} ) ) ]$. Taking only the dominant part, the complexity becomes $O (N_{out} N_{it} { (KNMJ)^2 ~ {\binom{M^J}{2}}} )$. It can be observed that the proposed Log-sum-exp scheme (\ref{maximinlogsumexp}) has reduced the computation complexity of SCMA CB design by a significant amount from DR  based CB  optimization problem  shown in (\ref{maximindro}).

\subsection{Theoretical BER}
SCMA is a multi-carrier scheme, i.e., the superimposed transmitted codeword $\textbf{s}_i= [s_i^1,\cdots, s_i^K]$ is a $K$ dimensional real codeword as discussed in Section \ref{sec_2b}.
In this section, the theoretical BER expression has been derived for the  IDGN incorporated SCMA-VLC system.  
 For only the thermal noise case, the PEP is given as (\ref{pepawgn}). However, in the case of shot noise, the overall variance changes and is dependent on $\varsigma^2$ and the intensity of the incident optical signal. 
The error rate for the superimposed codeword $\textbf{s}_i$ is 
\begin{align}
        P'(\textbf{s}_i )=
        \sum\limits_{\substack{i'=1 \\ i'\neq i}}^{M^J-1}
        P'(\textbf{s}_i \rightarrow \textbf{s}_{i'}) ~~ 
        \\=         \sum\limits_{\substack{i'=1 \\ i'\neq i}}^{M^J-1} 
        Q \left(  \sqrt{\sum_{k=1}^K \frac{ \norm{{s}_i^k-{s}_{i'}^k}^2}{2  \nu_{i,k}  }}\right),
\end{align}
where ${s}_i^k$ and ${s}_{i'}^k$, denotes the $k$th element of  the superimposed codeword $\textbf{s}_i$ and $\textbf{s}_{i'}$, respectively. The $k$th diagonal element of the covariance matrix $\Sigma_i$ of the superimposed codeword $\textbf{s}_{i}$ is denoted by $\nu_{i,k}$. The diagonal element $\nu_{i,k}$ denotes the total variance at $k$th RE when   $\textbf{s}_i$ was transmitted. 
The total probability of error is given as \cite{starqam_anapaper_13}  
\begin{align}\label{perror}
        P_{err}= \frac{1}{J \text{log}_2{M}} \sum_{i=1}^{M^J} P_a(\textbf{s}_i) \sum\limits_{\substack{i'=1 \\ i'\neq i}}^{M^J-1} h_d(\textbf{s}_{i'},\textbf{s}_i) P'(\textbf{s}_i \rightarrow \textbf{s}_{i'}),
\end{align}
where $P_a(\textbf{s}_i)$ denotes the \textit{a priori} probability of $\textbf{s}_i$, $h_d(\textbf{s}_{i'},\textbf{s}_i)$ denotes the bit error when $\textbf{s}_{i'}$ is chosen over $\textbf{s}_i$ and $P'(\textbf{s}_i \rightarrow \textbf{s}_{i'})$ denotes the PEP between $\textbf{s}_i$ and $\textbf{s}_{i'}$. 
\begin{figure}[htbp]
\centering
\includegraphics[scale=0.6,trim=3.5cm 8cm 4cm 8cm,clip]{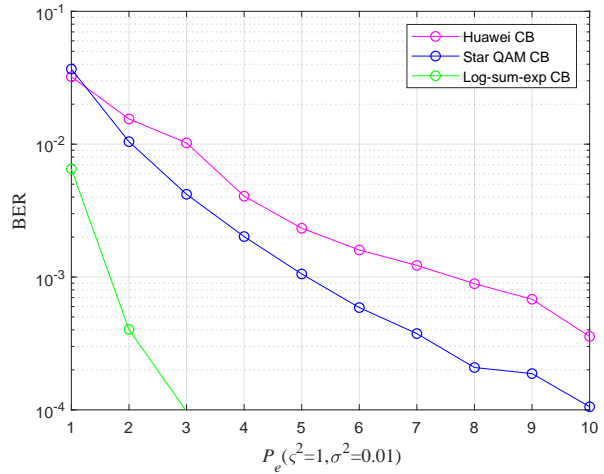}
\caption{BER performance comparison for $J=6$ between Star QAM based CB \cite{starqam}, Huawei CB \cite{scmapot_91}  and Log-sum-exp CBs for $ K=8,  \sigma^2=0.01, M=4$.}
\label{fig:rfvs_logsum_8scs}
\end{figure}

\begin{figure}[t]
\centering 
\subfigure[User 1]{\label{ex0}
{\includegraphics[width=0.15\textwidth,height=0.3\linewidth,trim=5cm 8cm 5cm 8cm,clip]{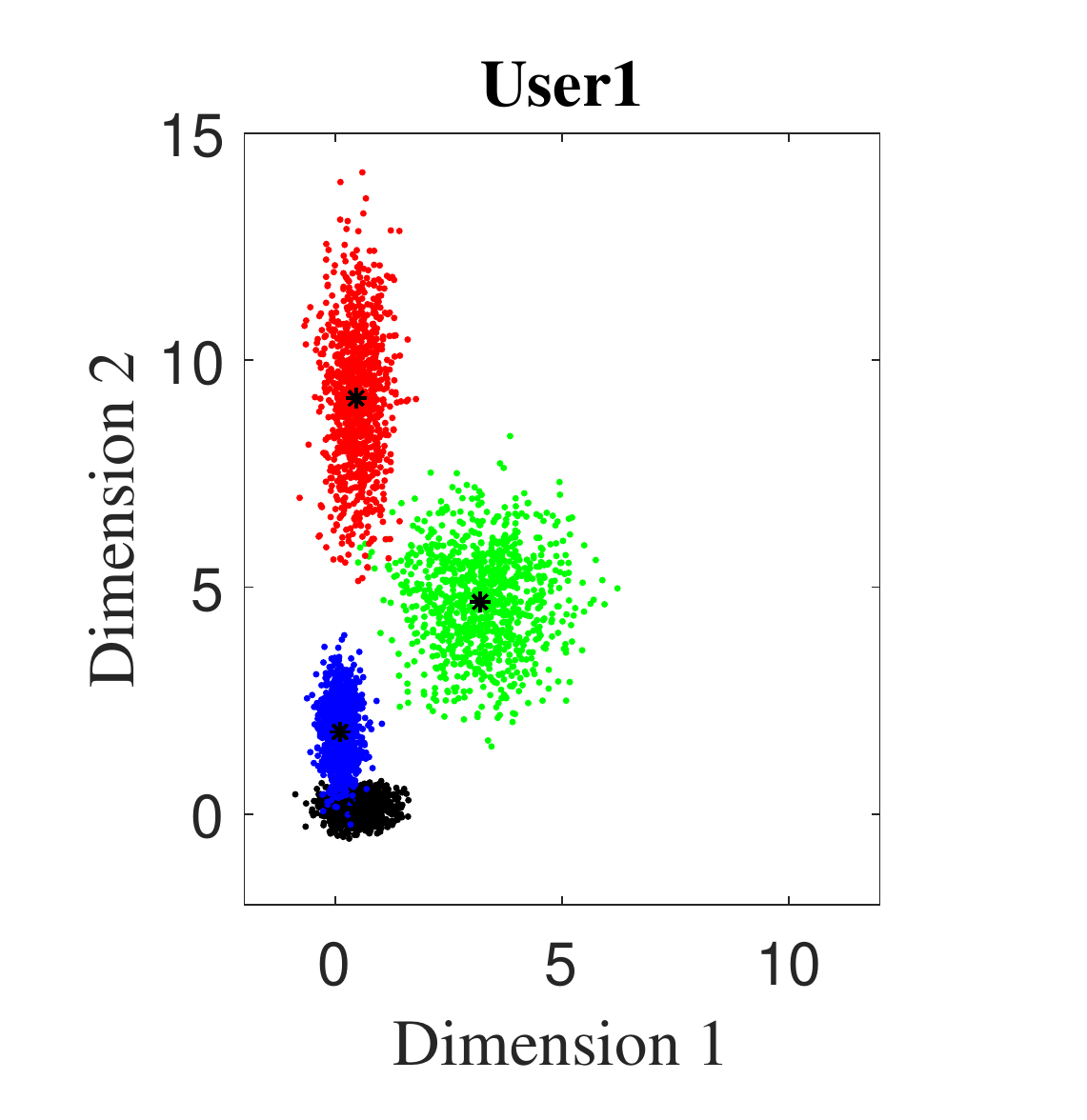}}}
\subfigure[User 2]{\label{ex1}
{\includegraphics[width=0.15\textwidth,height=0.3\linewidth,trim=5cm 8cm 5cm 8cm,clip]{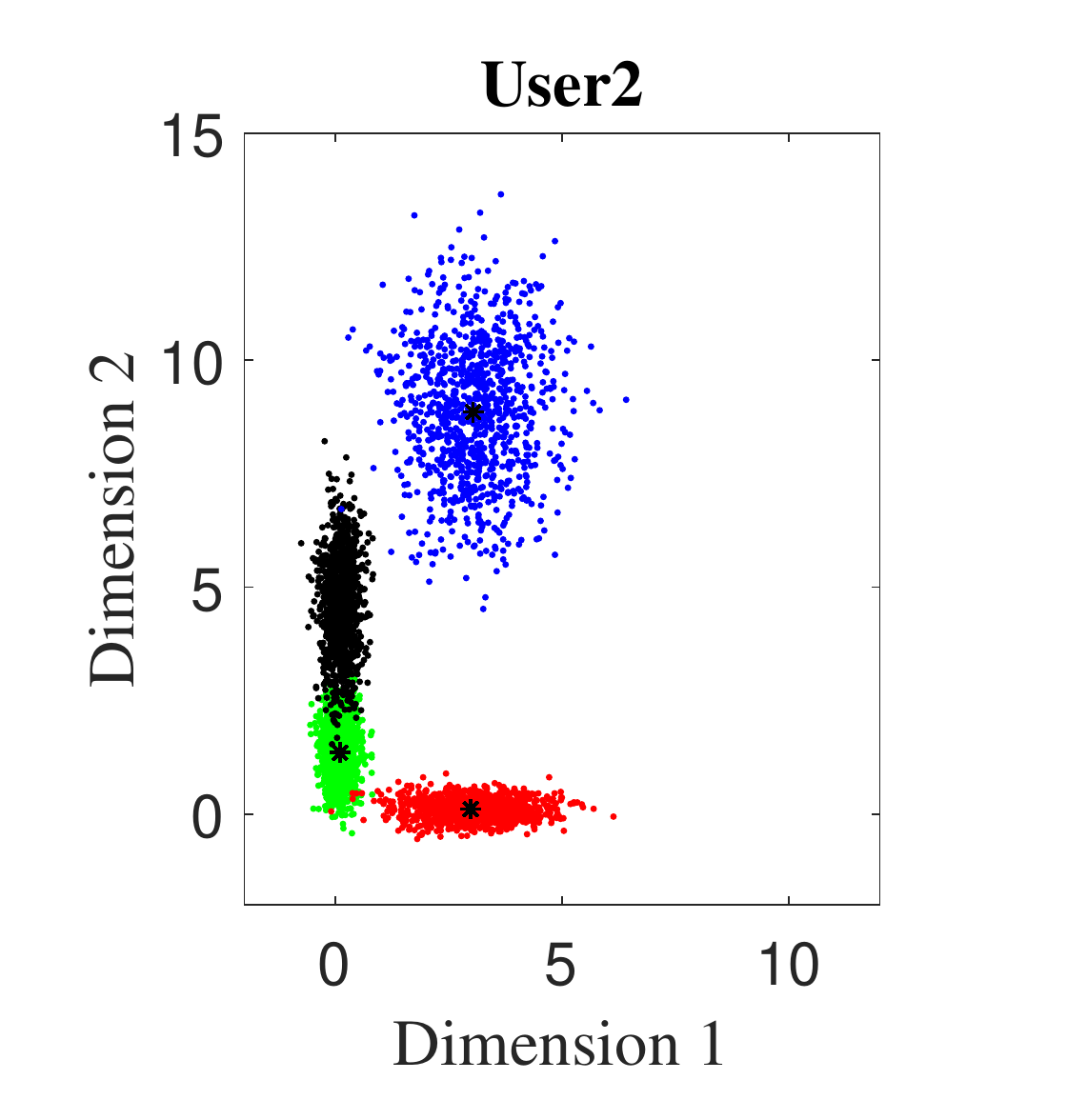}}}
\subfigure[User 3]{\label{ex3}
{\includegraphics[width=0.15\textwidth,height=0.3\linewidth,trim=5cm 8cm 5cm 8cm,clip]{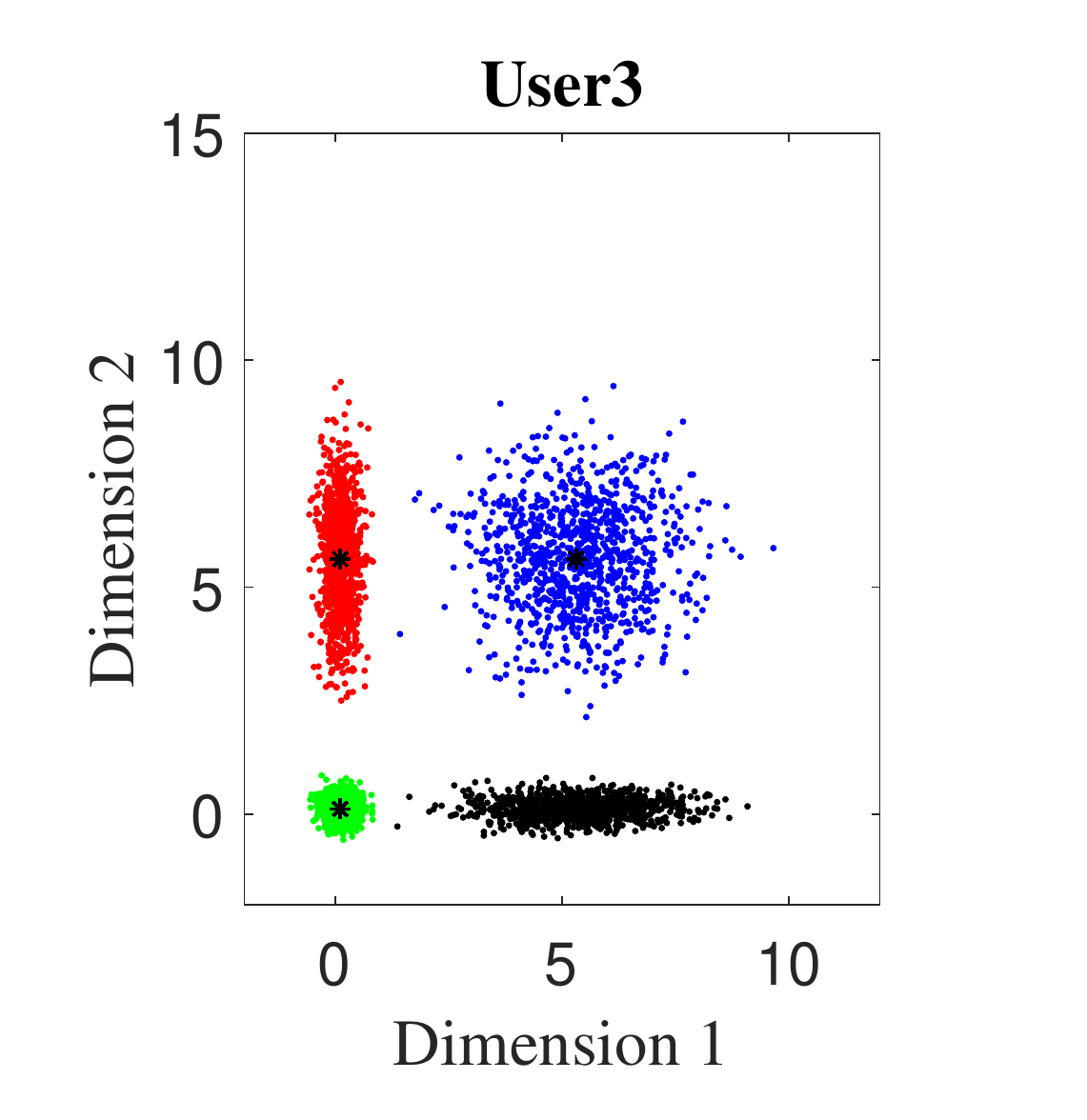}}}\\
\subfigure[User 1]{\label{ex4}
\includegraphics[scale=0.23,trim=5cm 8cm 5cm 8cm,clip]{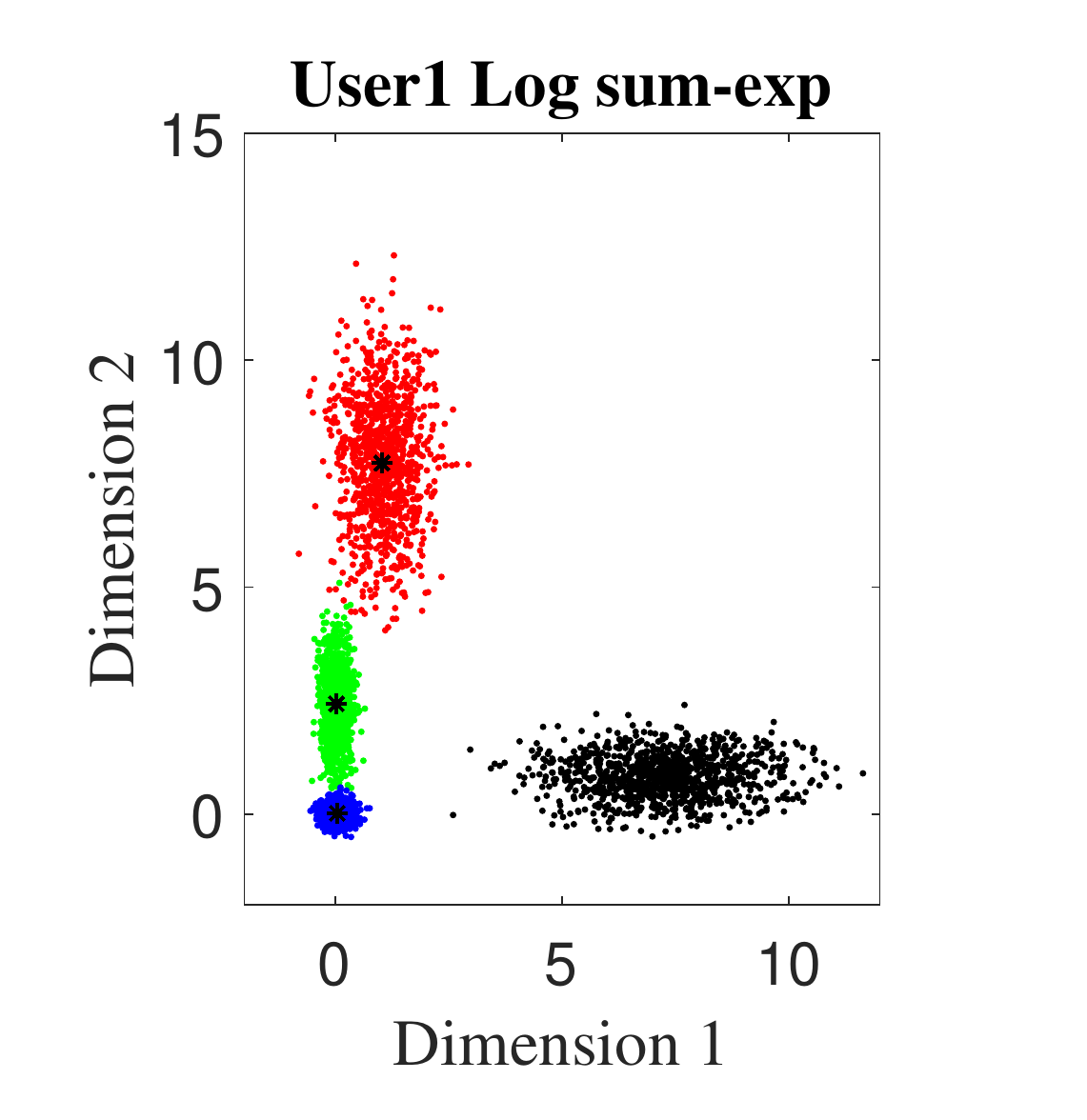}}\hfill
\subfigure[User 2]{\label{ex5}
\includegraphics[scale=0.23,trim=5cm 8cm 5cm 8cm,clip]{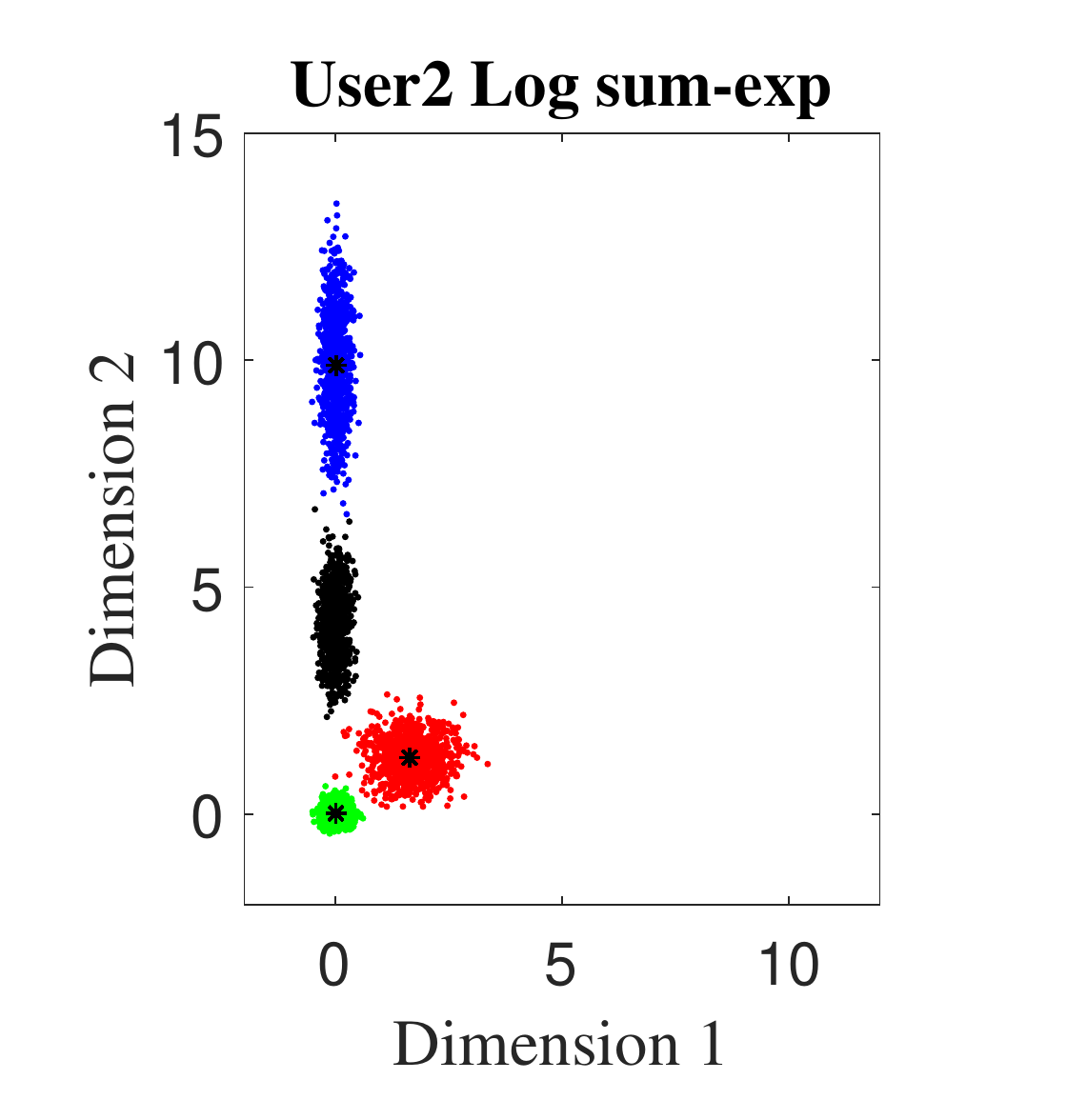}}\hfill
\subfigure[User 3]{\label{ex6}
\includegraphics[scale=0.23,trim=5cm 8cm 5cm 8cm,clip]{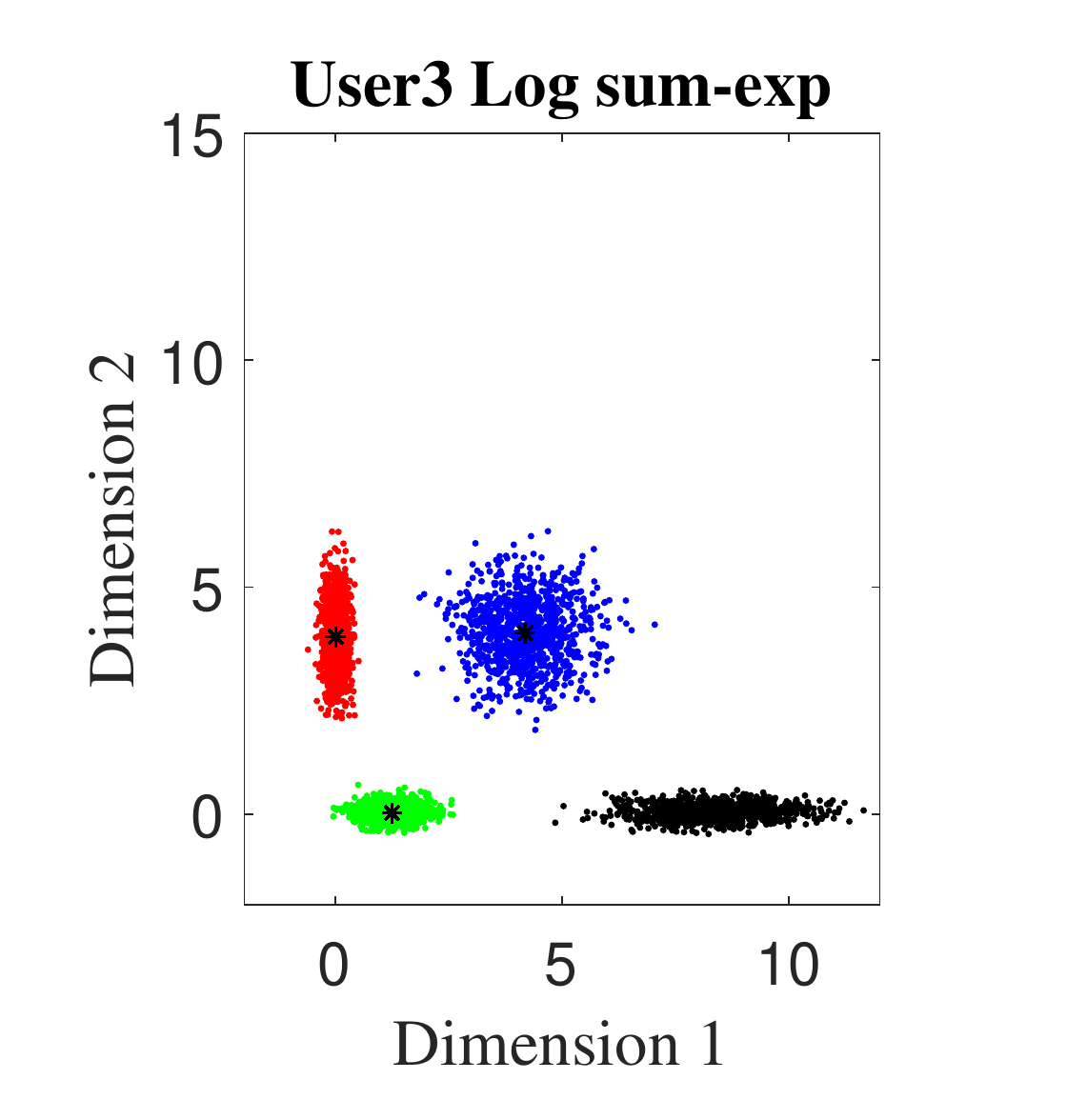}}
\caption{SCMA constellations scatter plots with $J=3,K=4,P_e=30, \varsigma^2=10$ (a)-(c): DR based constellation sets, (d)-(f): Log-sum-exp based constellation sets.}
\label{scatterplot_dro_logsum}
\end{figure}

\begin{figure}[htbp]
\centering
\includegraphics[scale=0.61,trim=3.5cm 8cm 4cm 8cm,clip]{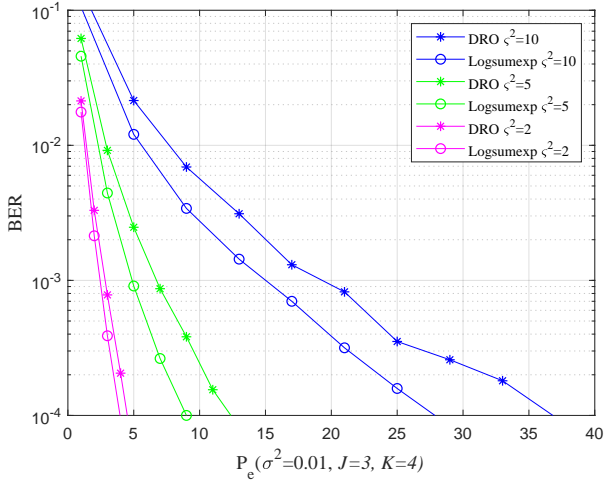}
\caption{BER performance comparison for $J=3$ between DR based CBs and Log-sum-exp CBs for $ \sigma^2=0.01,M=4$.}
\label{fig:dro_logsum3users}
\end{figure}

\section{Results and Discussion}
In this section, the simulation and theoretical results of the BER  performance of the proposed SCMA CB are presented.  Firstly,  using  \textbf{Algorithm  2},  we  can  generate the proposed CBs for different users. Further, Max-Log-MPA (as shown in \textbf{Algorithm 1}) has been used for multi-user detection.

\subsection{ RF based CBs Performance in VLC system}
We consider the case where RF-based CBs are used for transmitting data over a VLC system. We consider the $4 \times 6$ SCMA CBs from \cite{starqam} and \cite{scmapot_91}. RF CBs cannot be employed as it is, as the VLC system has  the requirement of real and positive signal transmission. So,  the complex codeword elements are divided into real and imaginary parts, which are then DC biased to get real and positive codeword elements \cite{starqam}, \cite{scmapot_91}. These real and positive codewords are  transmitted  over different REs . This way, the data of six users is transmitted over eight REs. Further, we have generated the Log-sum-exp based CB for six users and eight REs scenario using  \textbf{Algorithm  2}. Fig. \ref{fig:rfvs_logsum_8scs} shows the bit error rate (BER) performance of the Star QAM based CB \cite{starqam}, Huawei CB \cite{scmapot_91}  and Log-sum-exp CBs for $ K=8$ and $ \sigma^2=0.01$. It can be understood from Fig. \ref{fig:rfvs_logsum_8scs} that CBs designed solely for the VLC system (proposed Log-sum-exp based CB) outperforms the RF-based CBs for shot noise incorporated VLC system.   The performance gap observed in Fig. \ref{fig:rfvs_logsum_8scs} motivated the authors to design CB especially for the SCMA-VLC system. 

\subsection{The case of three users}
For the initial simulation, we consider the case where three users share four REs with $M=4$ and the number of non-zero elements in the codeword is $N=2$. The mapping matrices $\textbf{V}_1, \textbf{V}_2, \textbf{V}_3$ have the same values as shown in (\ref{mappingmatrix}) (i.e., $d_f=2$ for $k=1,2$ and $d_f=1$ for $k=3,4$).
For three users, the  CBs (DR based and Log-sum-exp based) are generated by solving the optimization problems (\ref{maximindro}, \ref{maximinlogsumexp}) (CBs are shown  in Appendix). For $J=3$, \textbf{Algorithm  2} is found to be converging for low values of $\beta$ (i.e., $\beta=10$).  
Fig. \ref{scatterplot_dro_logsum}. shows the 2-dimensional (2D) scatter plots of the generated DR based CBs  (\ref{maximindro}) and  the proposed Log-sum-exp based CBs  (\ref{maximinlogsumexp}) for $\varsigma^2=10$  and $P_e=30$.  It can be noted that the constellation points under DR criterion (\ref{maximindro}) are substantially overlapping as shown in Fig. \ref{ex0}-\ref{ex3}. However, the overlapping between proposed Log-sum-exp based constellation points for all three users is negligible as shown in Fig. \ref{ex4}-\ref{ex6}.
As the overlapping increases between different constellation points, it becomes difficult for the receiver to detect the symbols correctly, and thus, the overall performance deteriorates.  
\begin{figure}[t]
\centering
\subfigure[User 1]{\label{ex0e}
{\includegraphics[width=0.13\textwidth,height=0.3\linewidth,trim=3.25cm 8cm 4.5cm 8.25cm,clip]{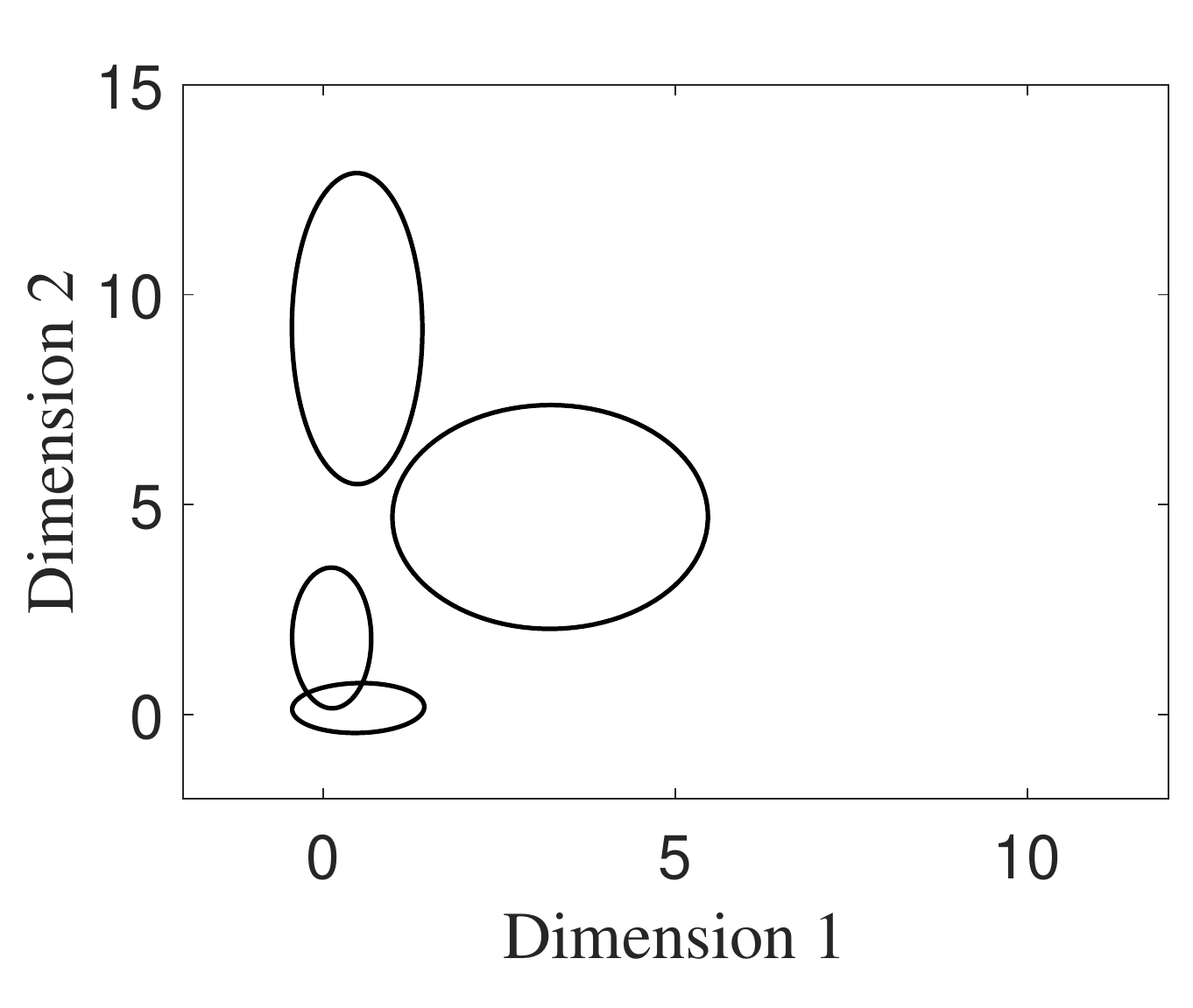}}}
\subfigure[User 2]{\label{ex1e}
{\includegraphics[width=0.13\textwidth,height=0.3\linewidth,trim=3.25cm 8cm 4.5cm 8.25cm,clip]{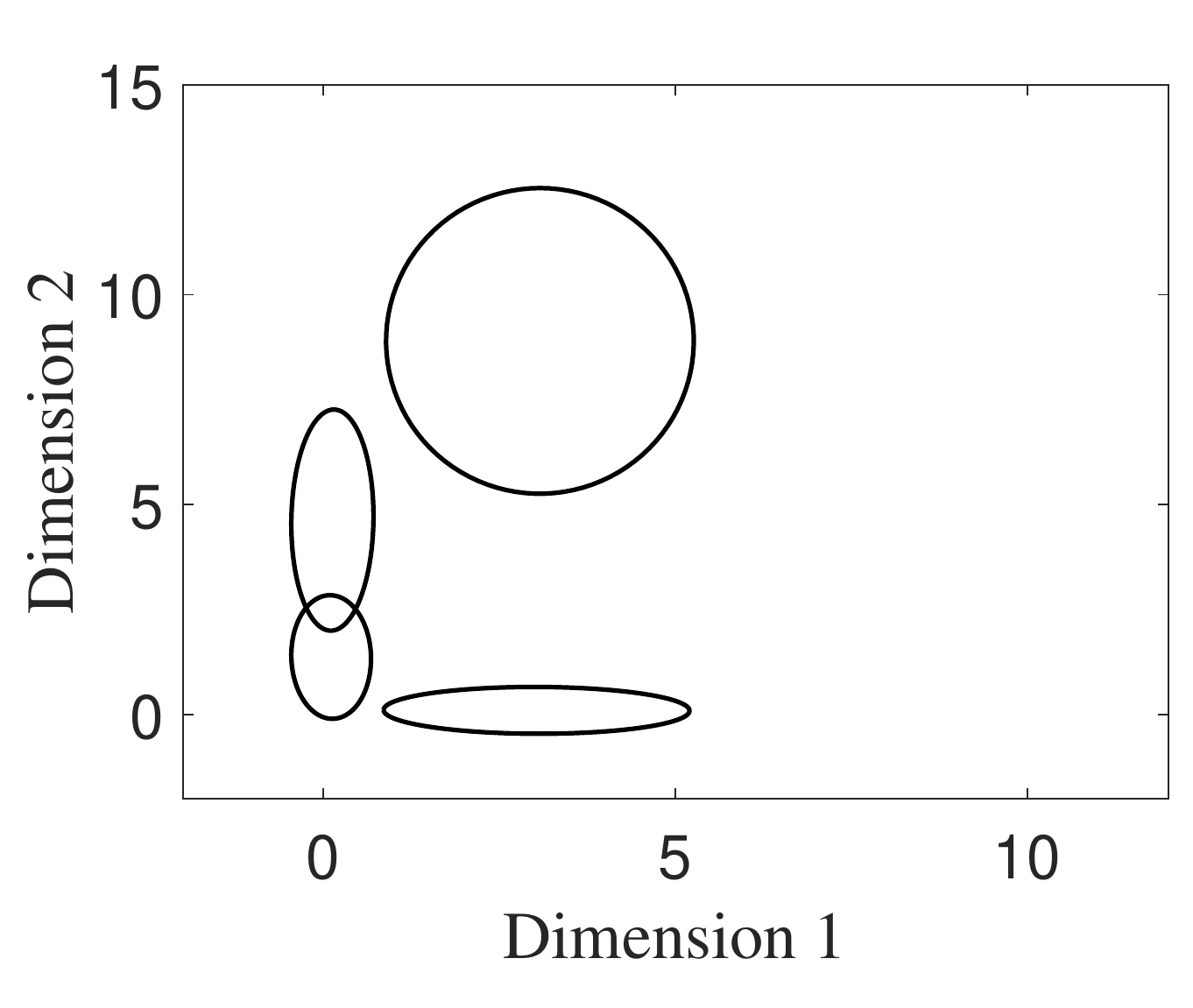}}}
\subfigure[User 3]{\label{ex3e}
{\includegraphics[width=0.13\textwidth,height=0.3\linewidth,trim=3.25cm 8cm 4.5cm 8.25cm,clip]{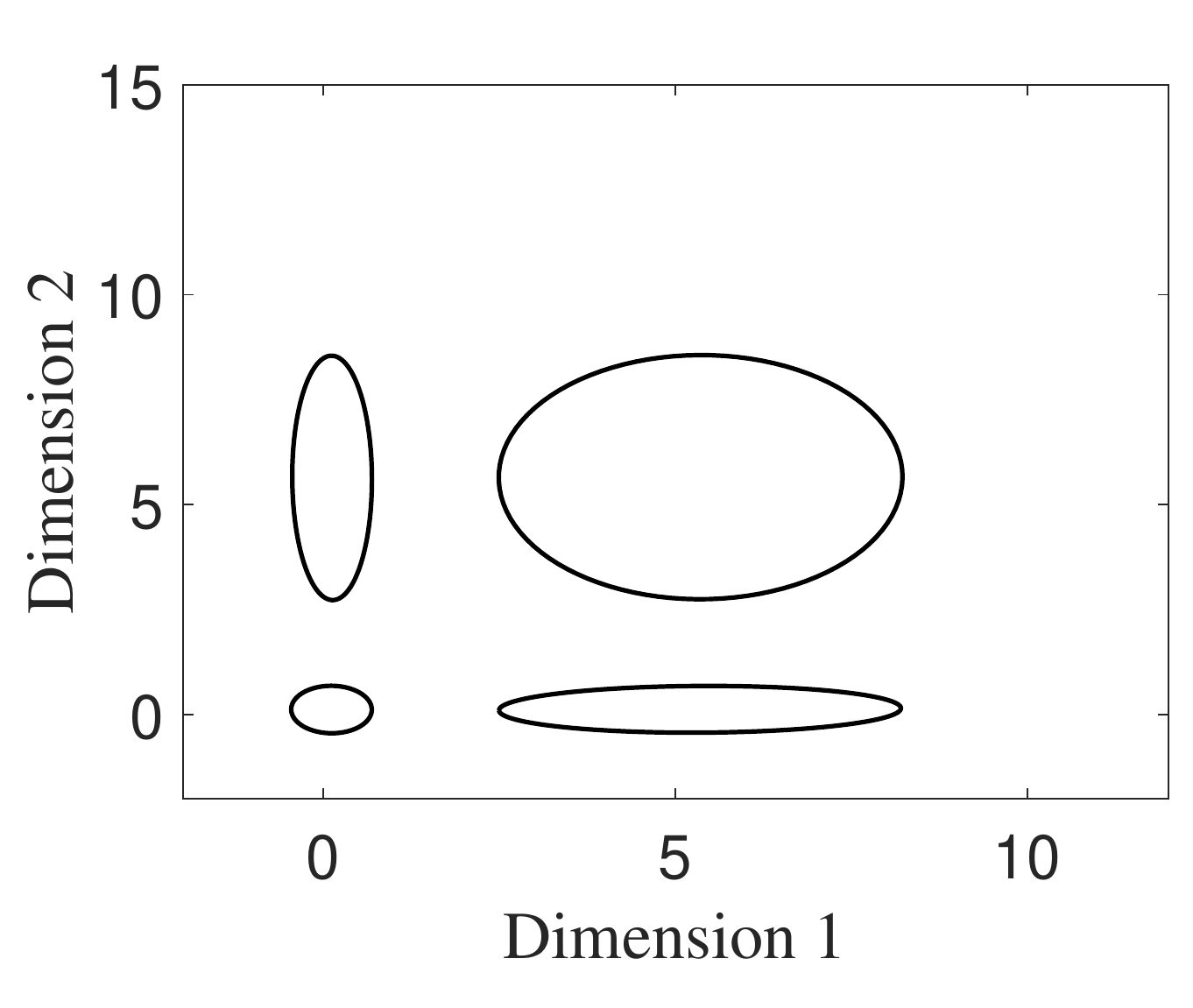}}}\\
\subfigure[User 1]{\label{ex4e}
\includegraphics[width=0.13\textwidth,height=0.3\linewidth,trim=3.25cm 8cm 4.5cm 8.25cm,clip]{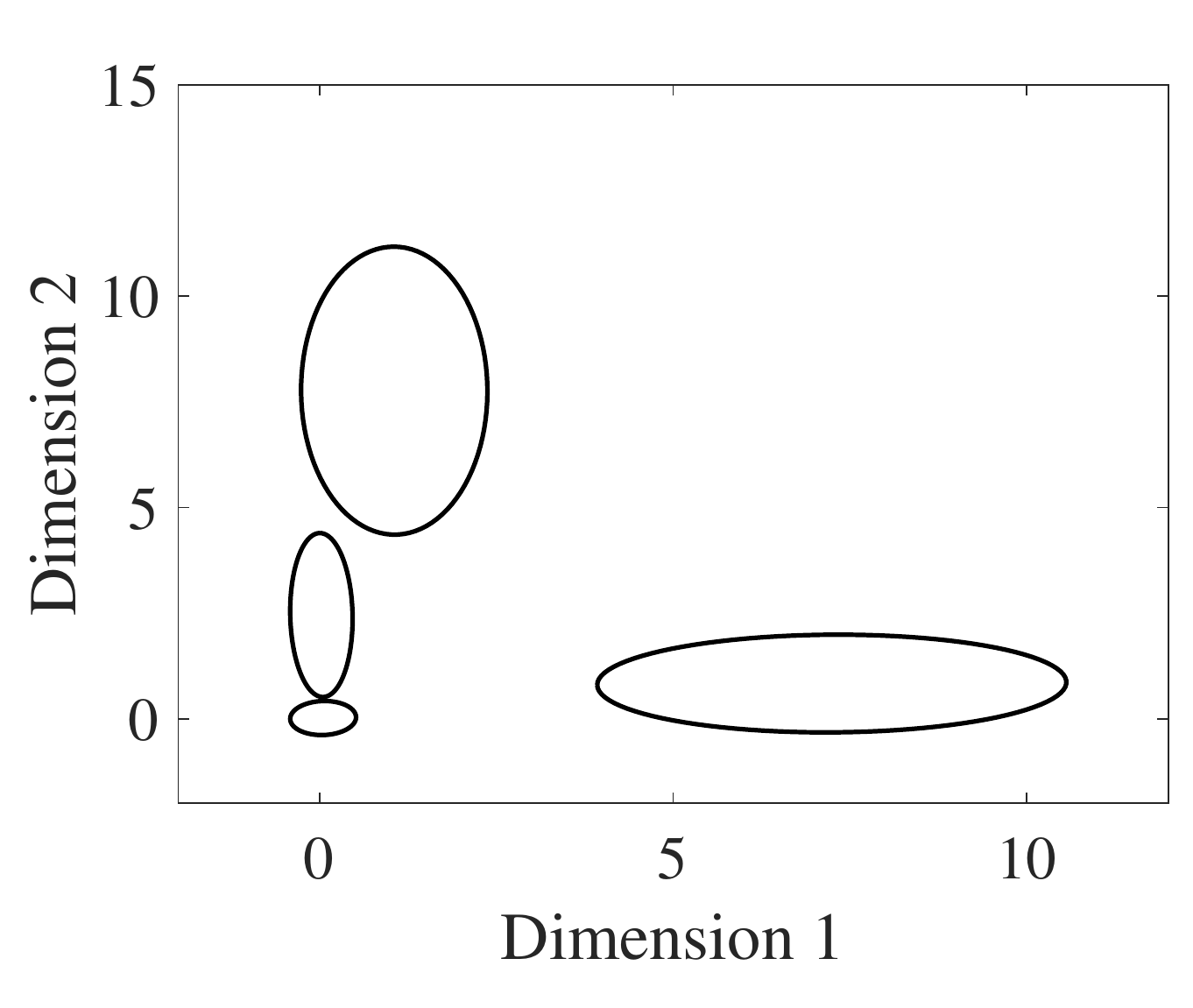} 
}
\subfigure[User 2]{\label{ex5e}
\includegraphics[width=0.13\textwidth,height=0.3\linewidth,trim=3.25cm 8cm 4.5cm 8.25cm,clip]{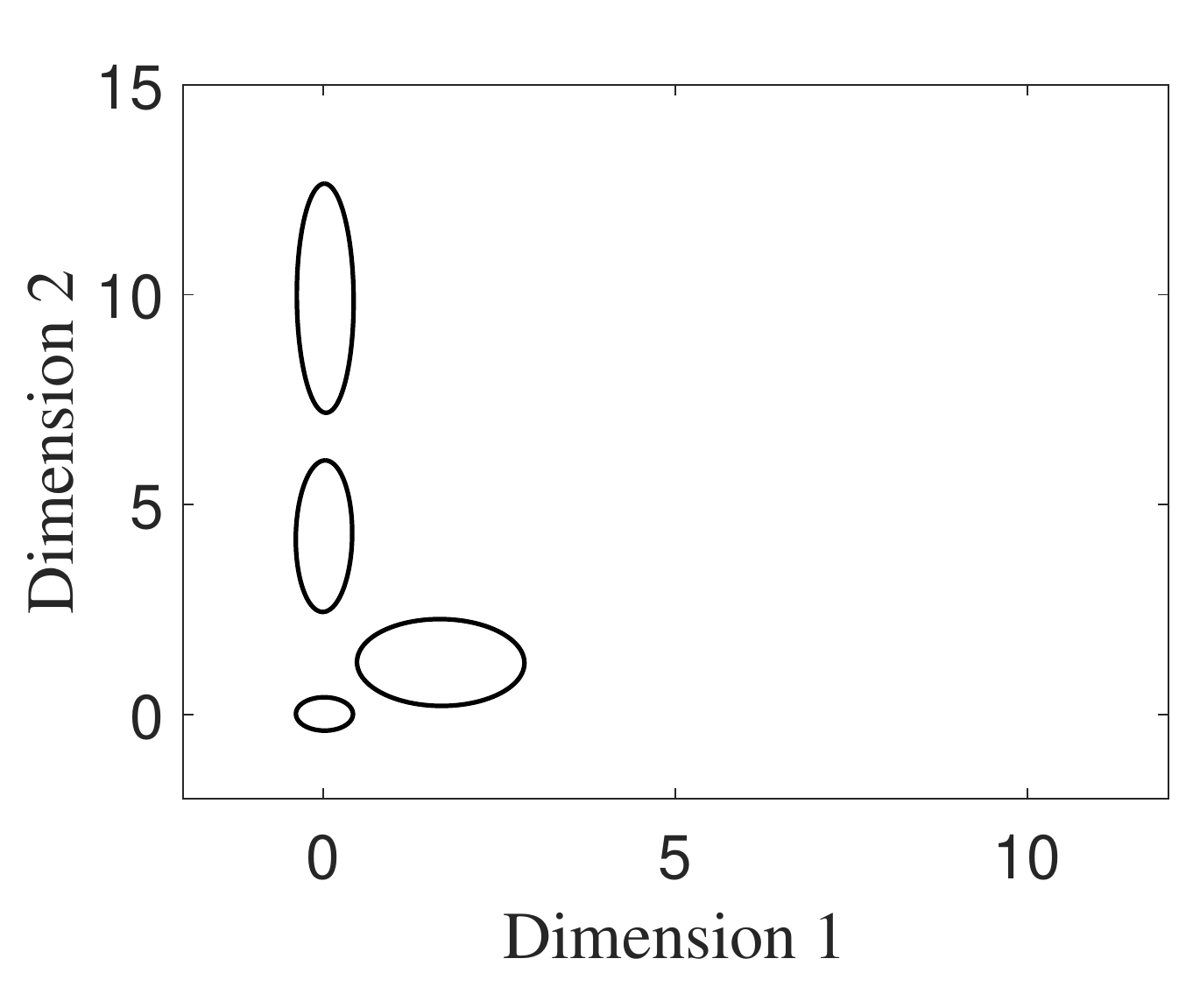}}
\subfigure[User 3]{\label{ex6e}
\includegraphics[width=0.13\textwidth,height=0.3\linewidth,trim=3.25cm 8cm 4.5cm 8.25cm,clip]{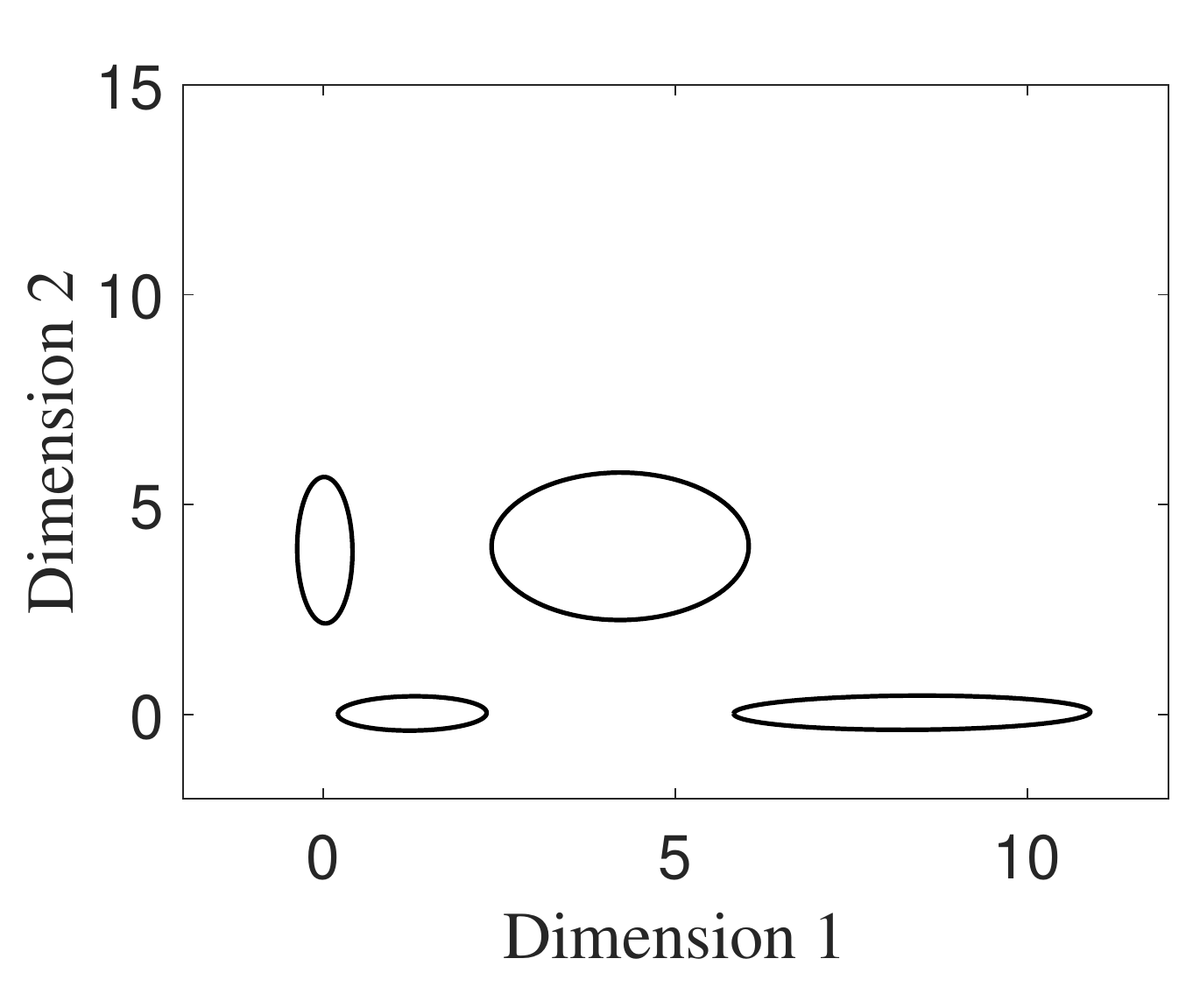}}
\caption{Lines of equal probability density functions of 2D SCMA constellations points with 95\% confidence interval  for $J=3,K=4,P_e=30, \varsigma^2=10$ (a)-(c): DR based EPDLs,  (d)-(f): Log-sum-exp based EPDLs.}
\label{epdl_3users}
\end{figure}

\begin{figure}[htbp]
\centering
\includegraphics[scale=0.61,trim=3.5cm 8cm 4cm 8cm,clip]{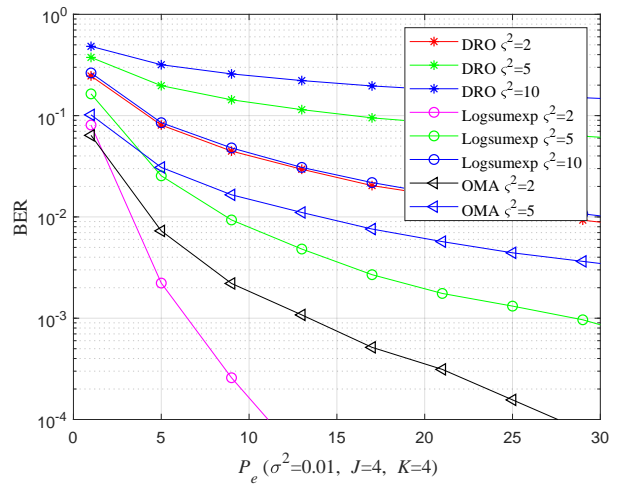}
\caption{BER performance comparison for $J=4$ between SCMA DR based CBs, Log-sum-exp CBs and OMA scheme for $ \sigma^2=0.01,M=4$.}
\label{fig:dro_logsumfourusers}
\end{figure}

\begin{figure}[htbp]
\centering
\subfigure[User 1]{\label{ex04}
{\includegraphics[scale=0.28,trim=5cm 8cm 5cm 8cm,clip]{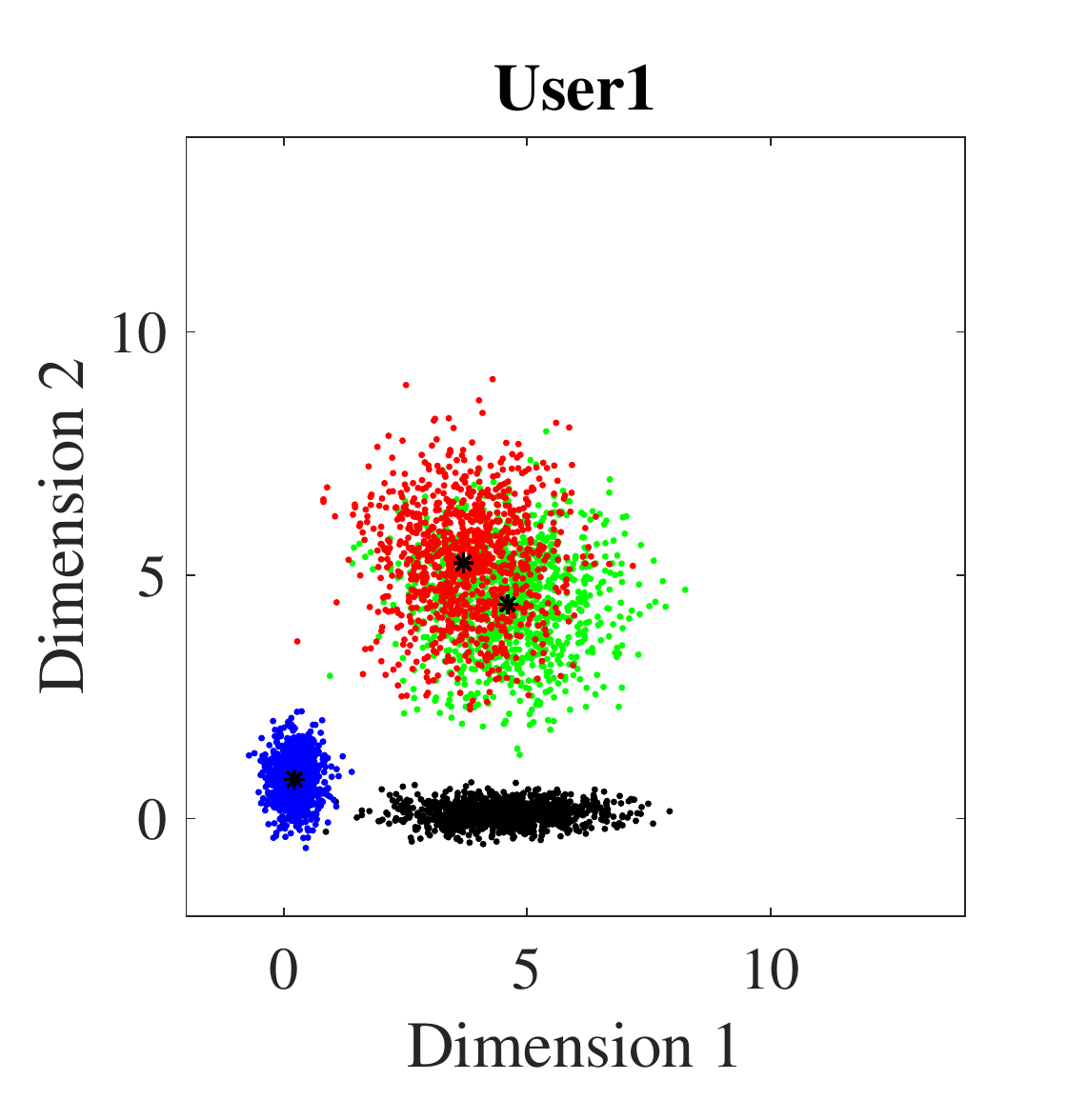}}}
\subfigure[User 2]{\label{ex14}
{\includegraphics[scale=0.28,trim=5cm 8cm 5cm 8cm,clip]{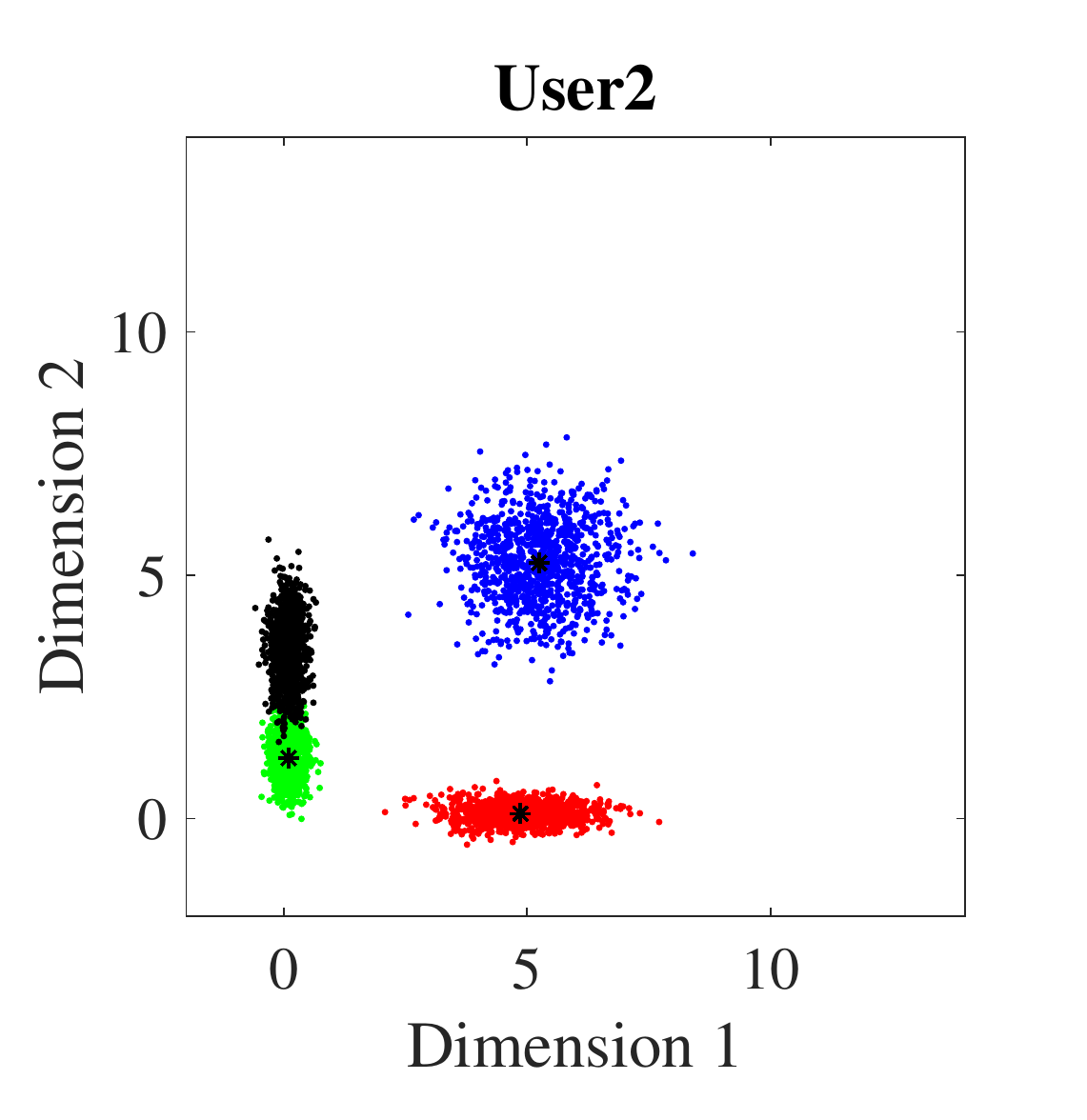}}}\\
\subfigure[User 3]{\label{ex34}
{\includegraphics[scale=0.28,trim=5cm 8cm 5cm 8cm,clip]{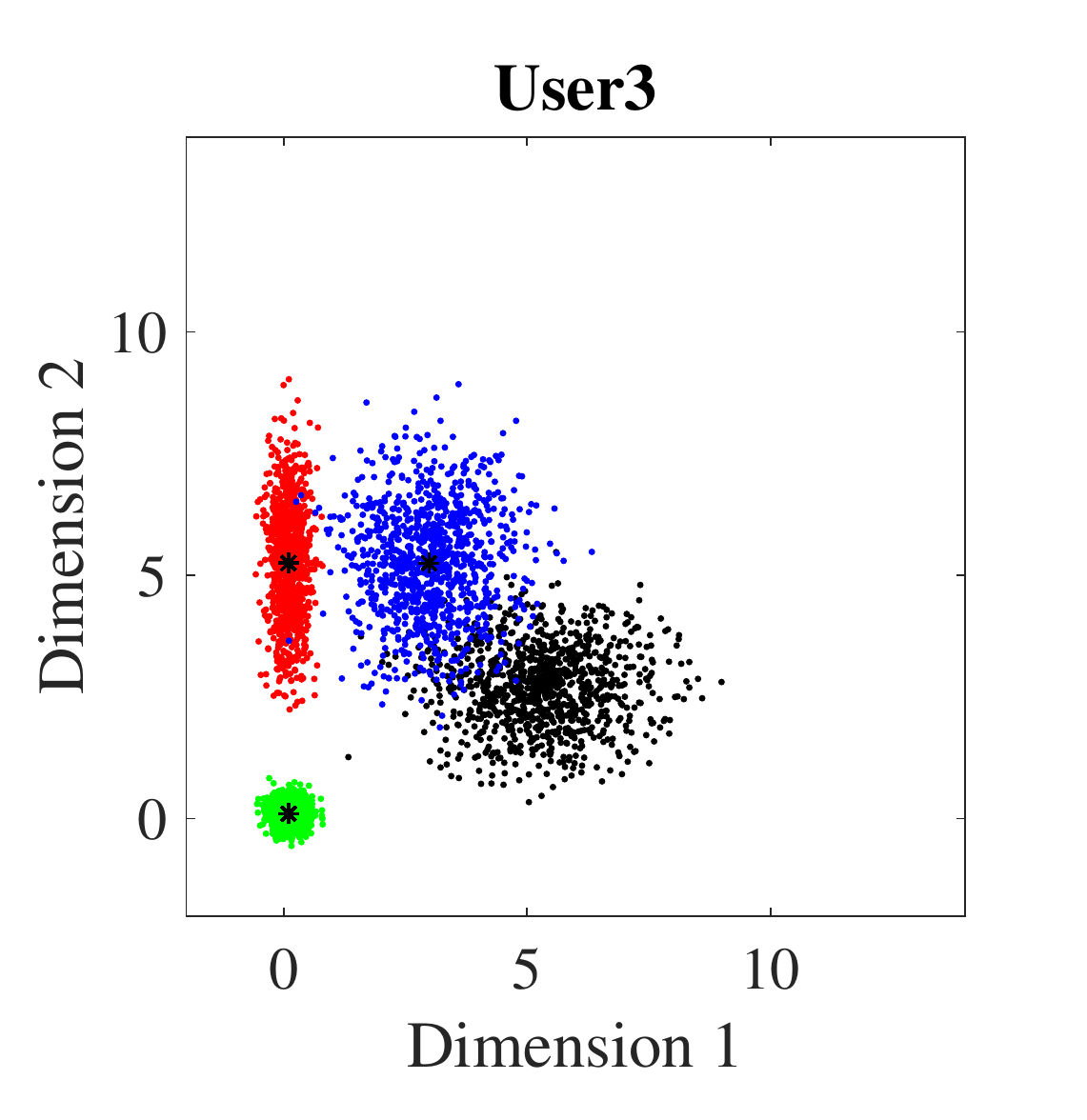}}}
\subfigure[User 4]{\label{ex44}
{\includegraphics[scale=0.28,trim=5cm 8cm 5cm 8cm,clip]{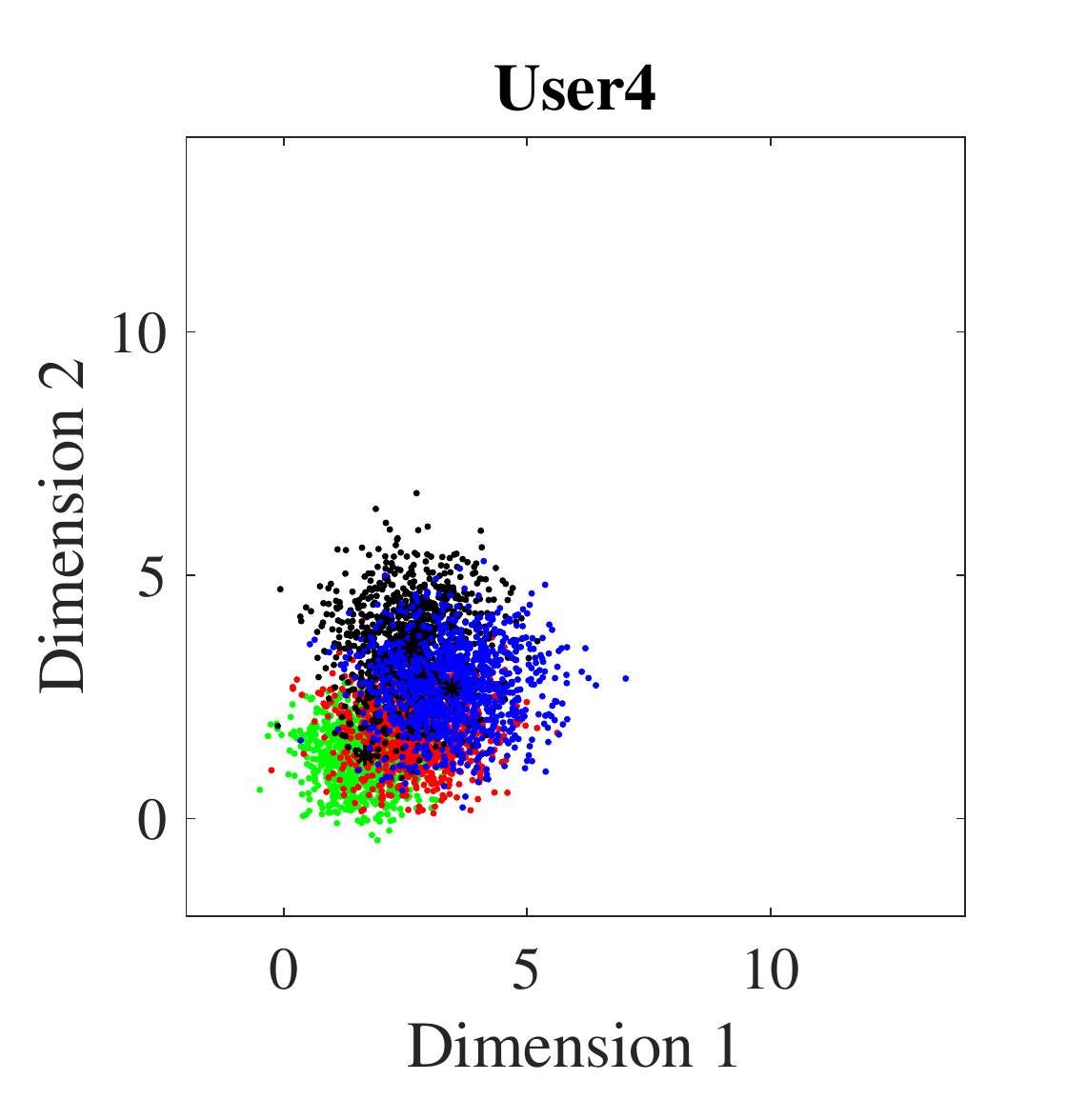}}}\\
\subfigure[User 1]{\label{ex54} {\includegraphics[scale=0.28,trim=5cm 8cm 5cm 8cm,clip]{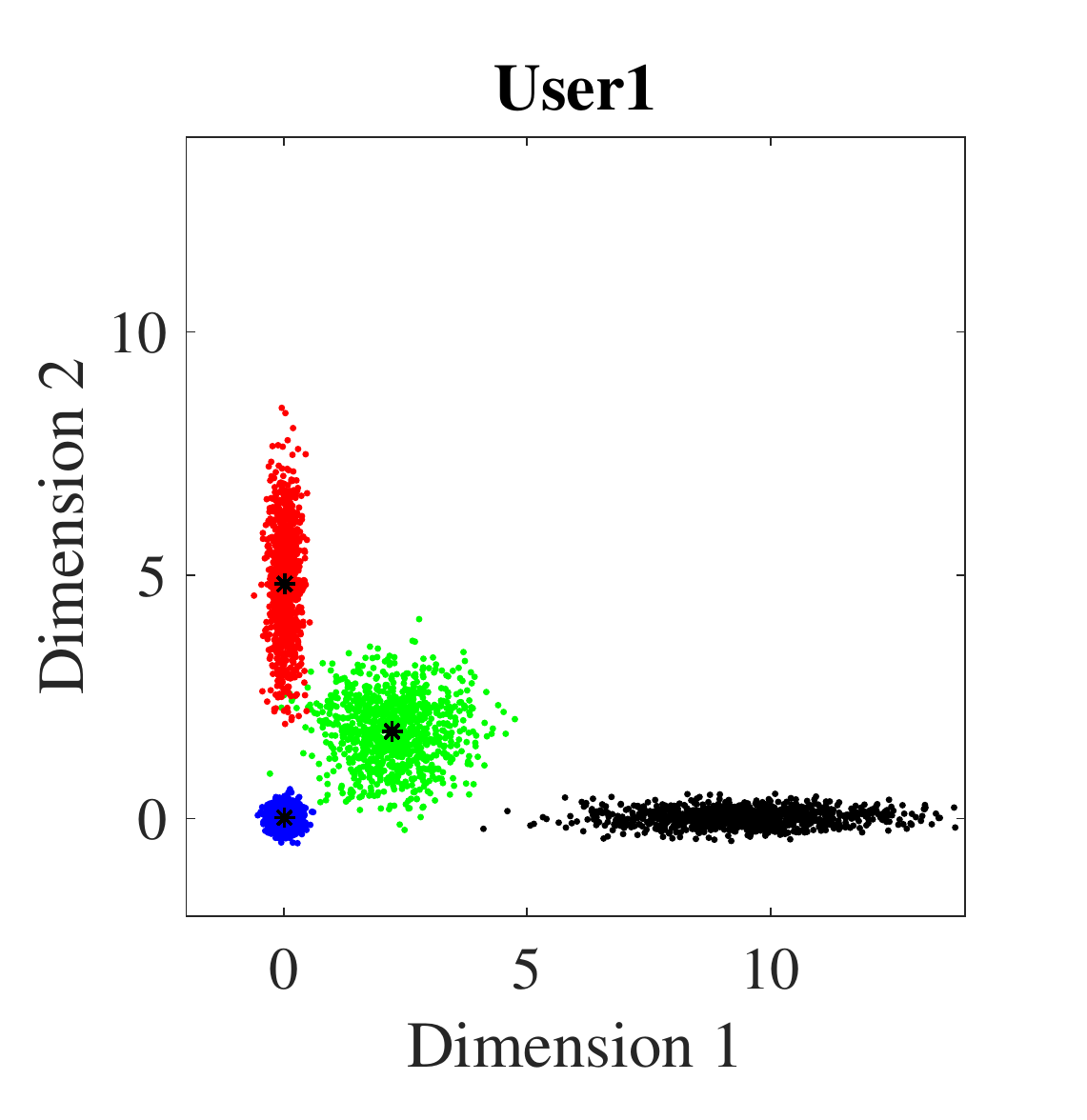}}}
\subfigure[User 2]{\label{ex64}
{\includegraphics[scale=0.28,trim=5cm 8cm 5cm 8cm,clip]{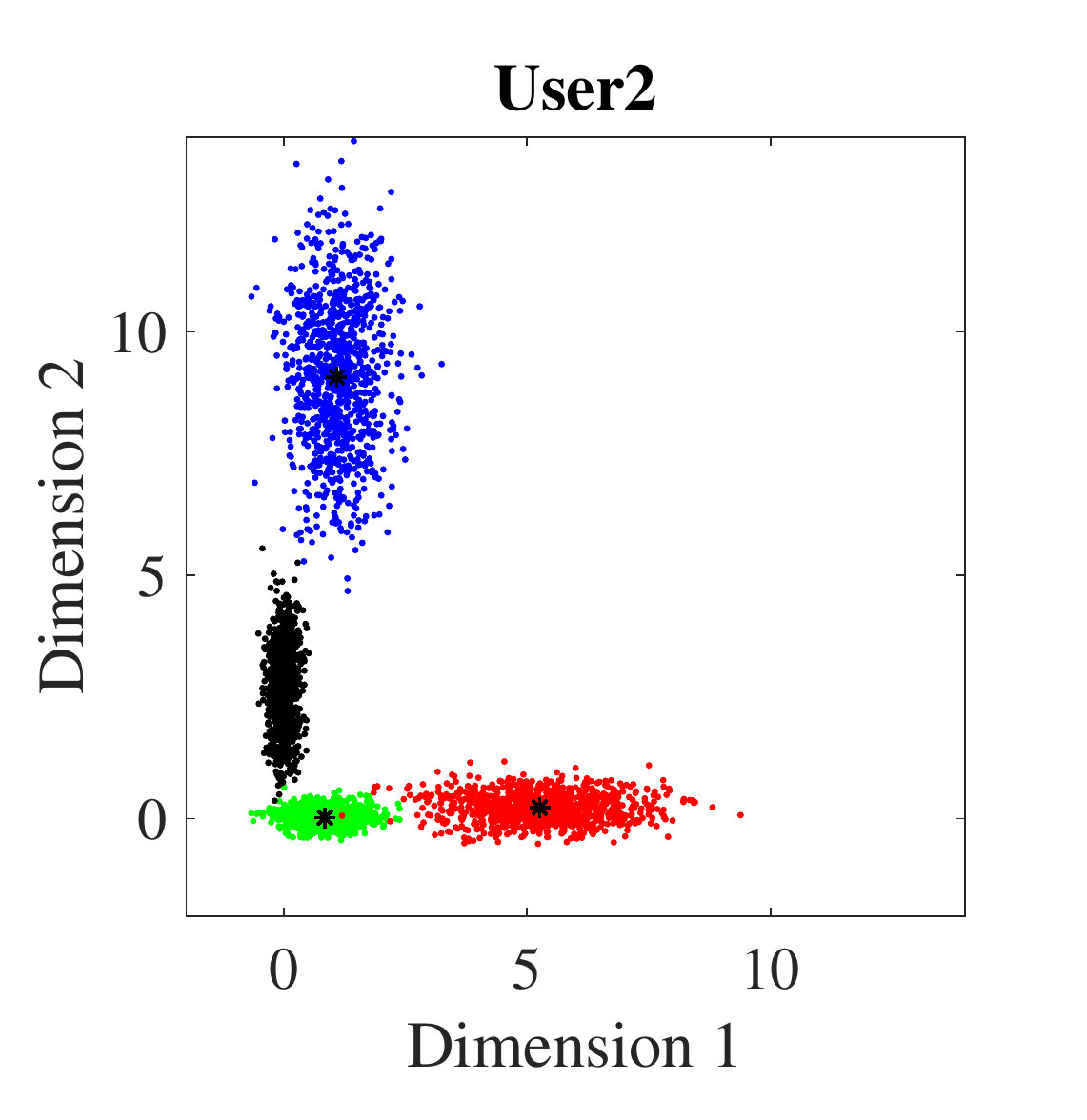}}}\\
\subfigure[User 3]{\label{ex74}
{\includegraphics[scale=0.28,trim=5cm 8cm 5cm 8cm,clip]{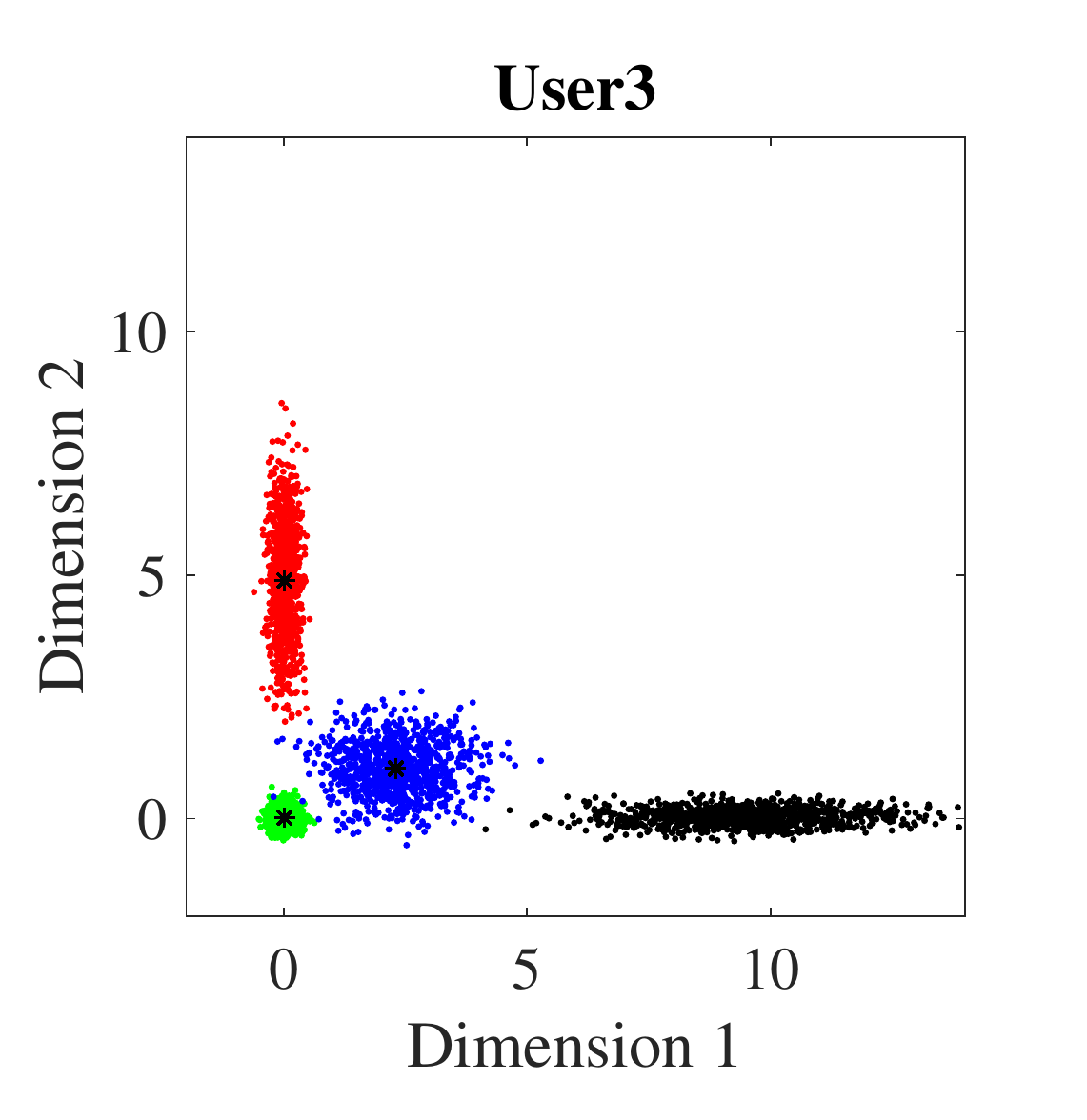}}}
\subfigure[User 4]{\label{ex84}
{\includegraphics[scale=0.28,trim=5cm 8cm 5cm 8cm,clip]{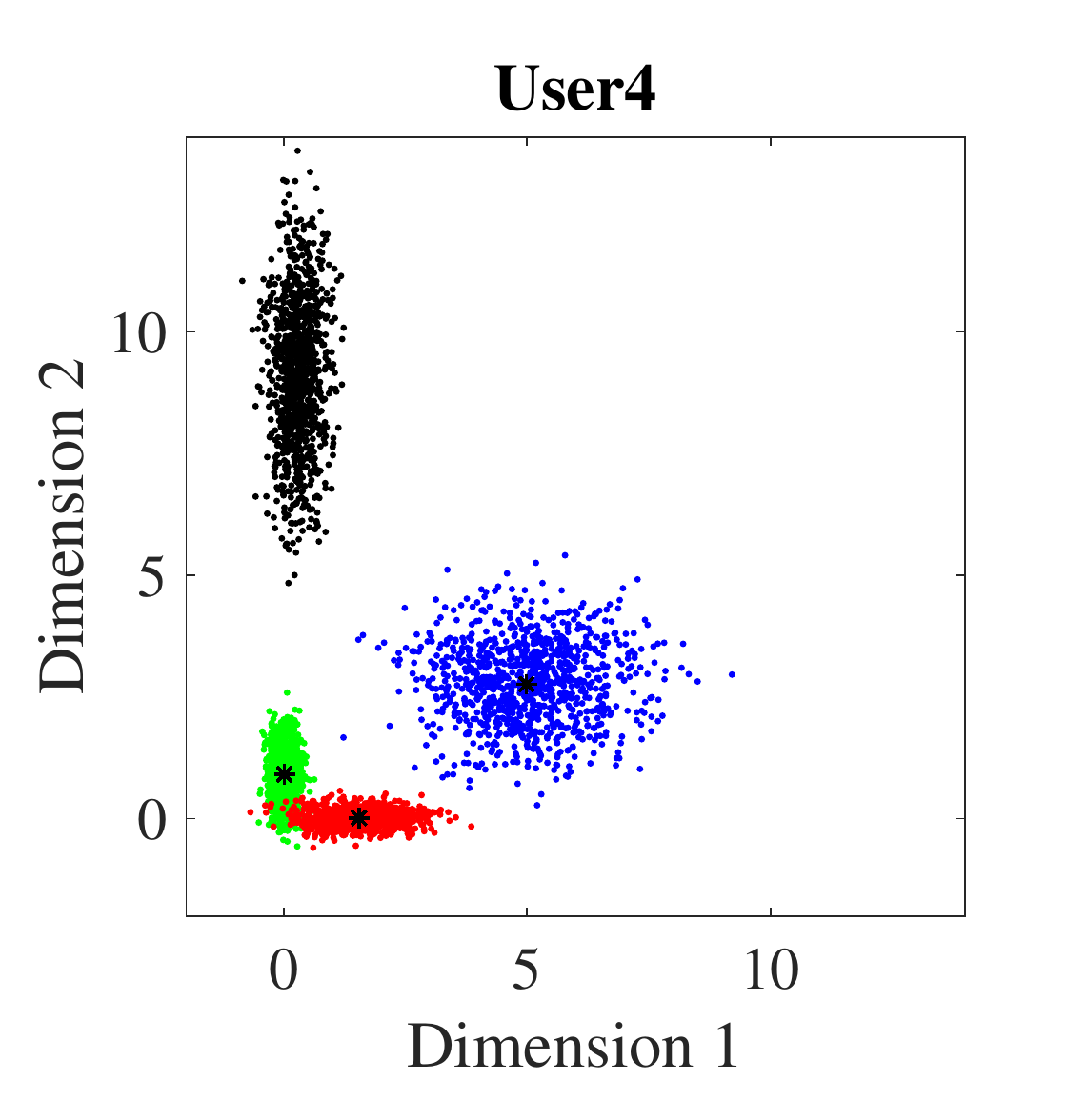}}}
\caption{SCMA constellations with $J=4,K=4,P_e=30, \varsigma^2=10$ (a)-(d): DR based constellation sets, (e)-(f): Log-sum-exp based constellation sets.}
\label{constlnsets_4users}
\end{figure}

Fig. \ref{fig:dro_logsum3users} shows the bit error rate (BER) performance for three users  of the proposed Log-sum-exp based (\ref{maximinlogsumexp}) SCMA CBs and the DR based CBs (\ref{maximindro}), respectively. The negligible overlapping in the case of log-sum-exp CBs as compared to DR CBs (in Fig. \ref{scatterplot_dro_logsum}), reasons the performance gain obtained in Fig. \ref{fig:dro_logsum3users}. For both CBs (DR based and Log-sum-exp based),  Max-Log-MPA (\textbf{Algorithm 1}) has been  used  for multi-user detection. It can be noticed that for the low value of $\varsigma^2$ (e.g., $\varsigma^2=2$), BER performance of both CBs is close, and for high values of $\varsigma^2$ (e.g., $\varsigma^2=5,10$), the proposed Log-sum-exp based CBs  yield better performance gain as compared to DR based CBs. \color{black}
{This suggests that the objective function with all the REDs  results in better CB (i.e., Log-sum-exp based CB) than the CB with only minimum and maximum values of REDs (i.e., DR based CB). } \color{black}
It is shown that our proposed CBs achieve the BER of $10^{-3}$ at $P_e=15$ while DR based CB achieve it at $P_e \approx 20$ for $\varsigma^2=10$.
\color{black}
{This suggests that for large values of $\varsigma^2$, the advantage of the proposed Log-sum-exp based CB is more significant than DR based CB.
Further, from the results it can inferred that,  we can reduce the dimming level to $\frac{30-4}{30}$ = 86.6\% of designed power ($ P_e=30$) for $\varsigma^2=2$.}
\color{black}

The bivariate normal density is specified by the 2D constellation points and the variance of these points in different dimensions. 
Fig. \ref{epdl_3users} shows the lines of the equal probability density (EPD) functions of the bivariate Gaussian.   The center of the ellipse is the {2D} constellation point and its shape is determined by the covariance matrix of the 2D constellation points. The principal axes of these ellipses are given by the eigenvectors of the covariance matrix and, semi-major and semi-minor axis are dependent on the eigenvalues, respectively \cite{dhs_chap2_epdl}.  For $N=2$, the $m$th constellation point of the $j$th user is $\textbf{c}_j^m \in \mathbb{R}_{>0}^{2 \times 1}$  and the covariance matrix of this 2D constellation point is denoted by $\mathbb{S}_{\textbf{c}_j^m}$. So, the center of the ellipse is $\textbf{c}_j^m$ and its shape is determined by  $\mathbb{S}_{\textbf{c}_j^m}$.
Fig. \ref{ex0e}-\ref{ex3e} shows the lines of EPD for DR based constellation points  and Fig. \ref{ex4e}-\ref{ex6e} shows lines of EPD for Log-sum-exp based constellation points,  respectively. 






\begin{figure}[htbp]
\centering
\includegraphics[scale=0.61,trim=4cm 8cm 4cm 8cm,clip]{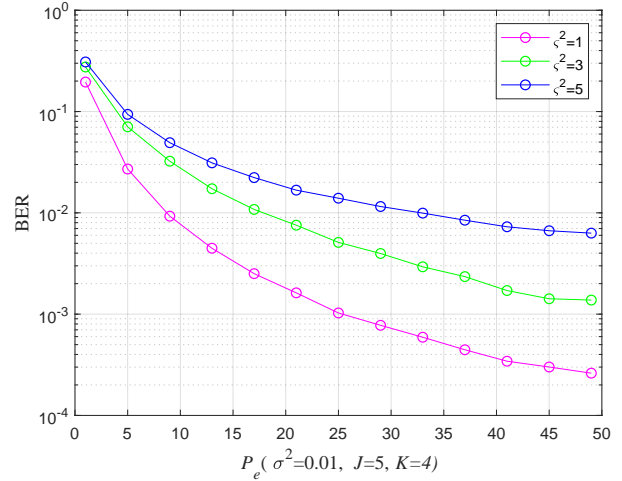}
\caption{BER performance comparison for $J=5$ based on Log-sum-exp CBs for $\sigma^2=0.01,M=4$.}
\label{fig:dro_logsumfiveusers}
\end{figure}

\begin{figure}[htbp]
\centering
\includegraphics[scale=0.61,trim=3.5cm 8cm 4cm 8cm,clip]{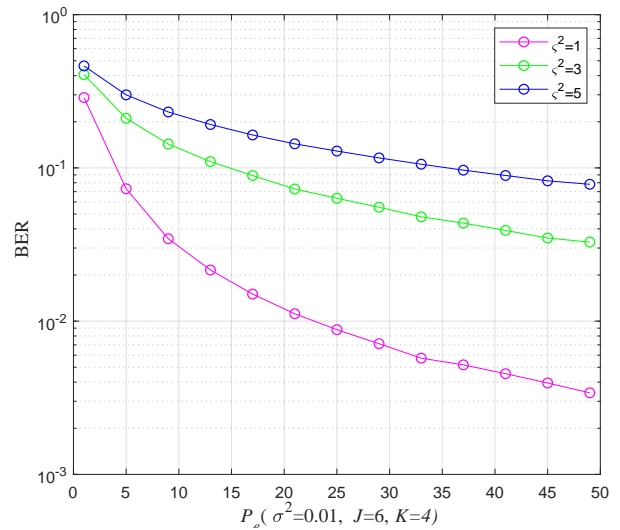}
\caption{BER performance comparison for $J=6$  Log-sum-exp CBs for $\sigma^2=0.01,M=4$.}
\label{fig:dro_logsumsixusers}
\end{figure}
\begin{figure}  
\centering
\subfigure[Three Users ]{\label{exthree}
{\includegraphics[scale=0.5,trim=3cm 8cm 4.5cm 8cm,clip]{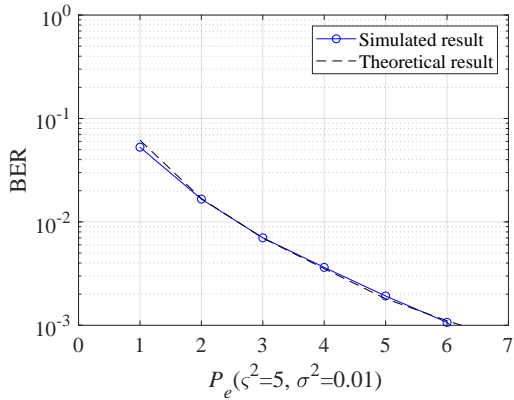}}}\\
\subfigure[Four Users]{\label{exfour}
{\includegraphics[scale=0.5,trim=3cm 8cm 4.5cm 8cm,clip]{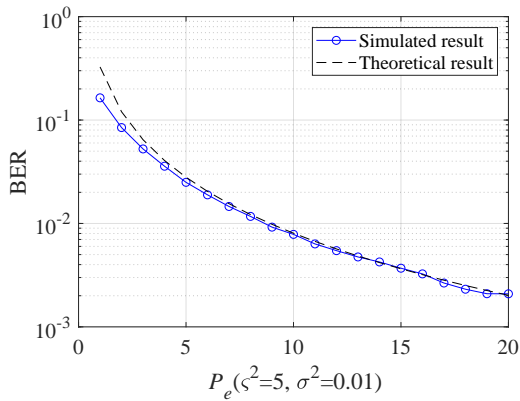}}}
\caption{Theoretical and Simulated BER performance comparison for (a)$J=3$ (b) $J=4$ for $\varsigma^2=5,\sigma^2=0.01$ and $M=4.$}
\label{anavssym_34users}
\end{figure}
\subsection{The case with four, five and six users ($J=4,5,6$)}

Consider the case where four users share four REs. \color{black}
{ The Log-sum-exp based CB is designed for four users using \textbf{Algorithm  2}, and its performance is compared with  DR based CB.} \color{black} Fig. \ref{fig:dro_logsumfourusers} shows the BER performance of four users where Log-sum-exp based CB outperforms DR based CBs for different values of $\varsigma^2$. The reason behind the performance difference can be understood from the level of overlapping between the constellation points for both CBs in Fig. \ref{constlnsets_4users}. It can be observed in Fig. \ref{constlnsets_4users} that the DR based 2D constellation points suffer from significant overlapping, especially for user 1 (Fig. \ref{ex04}) and user 4 (Fig. \ref{ex44}), while Log-sum-exp based 2D constellation points are barely overlapping. \color{black}
{This results in significant performance improvement for Log-sum-exp based CBs. Also in} Fig. \ref{fig:dro_logsumfourusers}, { the comparison with pulse amplitude modulation orthogonal multiple access (4 PAM-OMA) scheme is shown. It is  noted that the proposed algorithm outperforms the OMA scheme for different values of $\varsigma^2$.  This is expected as the OMA scheme offers no frequency diversity in contrast to the SCMA scheme.
    }
\color{black}

For five users, the complexity of generating DR based CB increases significantly. Further, the performance degrades drastically and requires a very high $P_e$ at BER of $10^{-3}$. Therefore, the effect of varying shot noise factor $\varsigma^2$ with only Log-sum-exp based CBs, when five users data is being transmitted over four REs ($\lambda=5/4)$ has been shown in Fig. \ref{fig:dro_logsumfiveusers}. The BER performance with respect to increasing $P_e$ is shown in Fig. \ref{fig:dro_logsumsixusers} for six users. Here, the effect of varying shot noise factor $\varsigma^2$ is shown when the proposed Log-sum-exp based CBs have been used for transmitting data over four REs.

In Fig. \ref{anavssym_34users}, the theoretical and simulated BER performances are compared. We find that the simulated curves match with the theoretical curves, especially for the high signal power region. These results verify the correctness of the simulated results  obtained in Fig. \ref{fig:dro_logsum3users} and Fig. \ref{fig:dro_logsumfourusers}. The results from Fig. \ref{fig:dro_logsum3users}, \ref{fig:dro_logsumfourusers},  \ref{fig:dro_logsumfiveusers} and \ref{fig:dro_logsumsixusers} have been summarized in Table \ref{table:2}. From Table \ref{table:2}, we have the electrical power required  to achieve BER of $10^{-3}$ for different users and VLC systems with different shot noise factors.
\rowcolors{2}{gray!10}{gray!40}
\begin{table}[htbp]
\caption{Electrical power required to get BER of $10^{-3}$, for different users and $\varsigma^2$ at $K=4$.}
\centering
\begin{tabular}{||c| c| c ||} 
 
 \hline
Number of    Users & $\varsigma^2$  & $P_e$ \\ 
 \hline\hline
\color{black} 3& \color{black}1&\color{black} 3 \\
 3& 5  & 5 \\
 \color{black}3&\color{black}10&\color{black} 15\\
 \color{black}4& \color{black}1& \color{black}6.5\\
 4& 5 & 29  \\
 5& 1&   25 \\
 \color{black}5 & \color{black}3 &\color{black} $>$50\\
 6 &1& $>$50\\
 \hline
\end{tabular}
\label{table:2}
\end{table} 
\section{Conclusion}
In this paper, a novel low-complex CB design technique has been proposed for SCMA-VLC system impaired by shot noise. This work emphasizes that the designing of specific CBs is required for the SCMA-VLC system. The proposed CBs are generated by optimizing the Log-sum of exponentials of all REDs between the superimposed codewords. With the inclusion of shot noise in the SCMA system, the overall noise variance is dependent on the strength of the incident signal, as shown in \textbf{Algorithm 1}. The proposed CBs provide improved BER performance as compared to CBs proposed in the existing literature along with reduced complexity. Further, the  theoretical analysis of BER is also performed for the shot noise incorporated SCMA-VLC system. The simulated results were validated, and both the theoretical and simulated curves are in good agreement for high signal power regions. \color{black}
{In the future, the performance of SCMA codebook can further be enhanced for lower $P_e$ by designing CBs considering lower dimming requirements,
and improve the performance of higher load factor scenarios in the SCMA-VLC system.} \color{black}

\begin{appendix}
The DR criteria based SCMA CB for VLC system obtained for three users for $\varsigma^2=5$ and $P_e=30$ is shown below:
\begin{equation*}
\begin{split}
    \textbf{CB}_1^{DR} =
\begin{bmatrix}
     0    &     0        & 0    &     0    \\
    3.1908  &  0.4595  &  0.1090    & 0.4907 \\
         0    &     0  &       0  &       0 \\
   4.6727  &  9.1620    &1.8083&    0.1476 \\
\end{bmatrix}, 
\end{split}
\end{equation*}
\begin{equation*}
    \textbf{CB}_2^{DR} =
\begin{bmatrix}
    0.1074  &  2.9890&    3.0405&    0.1300\\
         0  &       0   &      0   &      0\\
    1.3517 &   0.1077 &   8.8601    &4.5986\\
         0    &     0    &     0  &     0\\
\end{bmatrix}, 
\end{equation*}
\begin{equation*}
    \textbf{CB}_3^{DR} =
\begin{bmatrix}
    0.1119  &  0.1105&    5.3128&    5.3146\\
    0.1123   & 5.6158 &   5.6235&    0.1178\\
         0    &     0    &     0     &    0\\
         0     &    0     &    0      &   0\\
\end{bmatrix}. 
\end{equation*}
\end{appendix}
      The Log-sum-exp based SCMA CB  for VLC system  obtained for three users for $\varsigma^2=5$ and $P_e=30$ is shown below:
      
      \begin{equation*}
\begin{split}
    \textbf{CB}_1^{LS} =
\begin{bmatrix}
     0    &     0        & 0    &     0    \\
    2.7712 &   0.0100   & 0.0100 &   2.6317 \\
         0    &     0  &       0  &       0 \\
    4.4089    &9.1626 &   1.4151   & 0.0100 \\
\end{bmatrix}, \\ \\
    \textbf{CB}_2^{LS} =
\begin{bmatrix}
    0.0100 &   2.7785  &  2.5709  &  0.0100\\
         0  &       0   &      0   &      0\\
    0.0100   & 1.4036   & 9.1888 &    4.3893\\
         0    &     0    &     0  &     0\\
\end{bmatrix}, \\ \\
    \textbf{CB}_3^{LS} =
\begin{bmatrix}
    0.0100  &  5.4749  &   0.0100  &  5.4747\\
    0.0100   & 5.4797   & 5.4796    &0.0100\\
         0    &     0    &     0     &    0\\
         0     &    0     &    0      &   0\\
\end{bmatrix}. \\
\end{split}
\end{equation*}

Next, the SCMA CB for VLC system  generated for four users based on Log-sum-exp for $\varsigma^2=5$ and $P_e=30$ is shown below:
\begin{equation*}
\begin{split}
    \textbf{CB}_1^{LS} =
\begin{bmatrix}
        0   &      0  &       0  &       0\\
    1.5383   & 3.0405  &  0.0167  &  8.9796\\
         0   &      0   &      0   &      0\\
    1.3426    &5.0483    &0.0173    &0.0374 \\
\end{bmatrix}, \\
\end{split}
\end{equation*}
\begin{equation*}
    \textbf{CB}_2^{LS} =
\begin{bmatrix}
    0.0191 &   3.5636  &  4.1652  &  0.9455\\
         0 &        0  &       0   &      0\\
    0.0156  &  0.5631  &  8.9683    &2.7765\\
         0   &      0  &       0     &    0\\
\end{bmatrix}, \\ \\
\end{equation*}
\begin{equation*}
    \textbf{CB}_3^{LS} =
\begin{bmatrix}
   0.0278  &  6.5059   & 0.0257   & 6.4581\\
    0.0296 &   4.1903   & 4.2063   & 0.0268\\
         0  &       0    &     0    &     0\\
         0   &      0     &    0     &    0\\
\end{bmatrix}. \\
\end{equation*}
\begin{equation*}
   \textbf{CB}_4^{LS} =
\begin{bmatrix}
     0   &      0    &     0     &    0\\
         0 &        0 &        0  &       0\\
    0.0207  &  1.5398  &  5.0348   & 0.0264\\
    0.0154   & 2.3810   & 0.3892    &9.2690\\
\end{bmatrix}. \\
\end{equation*}
Similarly, the SCMA CB for VLC system  based on the proposed scheme for five users for $\varsigma^2=5$ and $P_e=30$ is shown below:
      \begin{equation*}
\begin{split}
    \textbf{CB}_1^{LS} =
\begin{bmatrix}
       0  &       0   &      0  &       0 \\
    1.8229 &    0.8713&    0.0707&    9.3633\\
         0  &       0  &       0  &       0\\
    0.2044   & 4.4911   & 1.5872   & 0.5904 \\
\end{bmatrix}, 
\end{split}
\end{equation*}
\begin{equation*}
    \textbf{CB}_2^{LS} =
\begin{bmatrix}
   1.3717  &  4.8155   & 1.2190  &  0.1399\\
         0 &        0   &      0  &       0\\
    0.0725  &  1.1399   & 9.0214   & 2.4409\\
         0   &      0    &     0    &     0\\
\end{bmatrix}, 
\end{equation*}
\begin{equation*}
    \textbf{CB}_3^{LS} =
\begin{bmatrix}
  0.1790 &   0.3468  &  2.3449  &  8.1443\\
    0.0915&    2.9830 &   5.7936 &   0.3253\\
         0 &        0  &       0  &       0\\
         0  &       0   &      0   &      0\\
\end{bmatrix}, 
\end{equation*}
\begin{equation*}
   \textbf{CB}_4^{LS} =
\begin{bmatrix}
      0  &       0    &     0    &     0\\
         0&         0  &       0  &       0\\
    0.0936 &   1.8345   & 4.8909   & 0.6543\\
    1.1709  &  0.1005    &2.2914    &9.0194\\
\end{bmatrix}, 
\end{equation*}
\begin{equation*}
   \textbf{CB}_5^{LS} =
\begin{bmatrix}
      3.1438  &  0.1500   & 6.2322 &   1.5089\\
         0    &     0      &   0    &     0\\
         0     &    0       &  0     &    0\\
    0.1397   & 1.4186  &  3.3720&    6.4569
\end{bmatrix}. 
\end{equation*}
At last, we show the SCMA CB for VLC system  based on the proposed scheme for six users for $\varsigma^2=5$ and $P_e=30$ as follows:
\begin{equation*}
\begin{split}
    \textbf{CB}_1^{LS} =
\begin{bmatrix}
          0   &      0    &     0 &        0\\
    0.8677    &2.1631   & 0.0295   & 6.5262\\
         0     &    0    &     0    &     0\\
    1.8603   & 8.1426   & 0.0594&    0.0741 \\
\end{bmatrix}, 
\end{split}
\end{equation*}
\begin{equation*}
    \textbf{CB}_2^{LS} =
\begin{bmatrix}
  0.8433 &   5.9356  &  2.0682  &  0.0335\\
         0&         0 &        0 &        0\\
    0.0379 &   0.0580  &  8.6673  &  1.8618\\
         0  &       0   &      0   &      0\\
\end{bmatrix}, 
\end{equation*}
\begin{equation*}
    \textbf{CB}_3^{LS} =
\begin{bmatrix}
   0.1106  &  0.0293  &   1.9845  &  9.2333\\
    0.0687  &  4.7442  &  2.2192  &  0.0930\\
         0   &      0   &      0   &      0\\
         0    &     0    &     0    &     0\\
\end{bmatrix}, 
\end{equation*}
\begin{equation*}
   \textbf{CB}_4^{LS} =
\begin{bmatrix}
       0   &      0    &     0     &    0 \\
         0  &       0   &      0    &     0\\
    0.0476   & 1.5733    &5.8671    &0.0408\\
    1.0104   & 0.0347    &0.4297    &8.9561\\
\end{bmatrix}, \end{equation*}
\begin{equation*}
   \textbf{CB}_5^{LS} =
\begin{bmatrix}
       3.2891  &  0.0679  &  3.8873 &   0.1111\\
         0     &    0      &   0     &    0\\
         0      &   0       &  0      &   0\\
    0.0304   & 1.3519   & 3.6688&    5.7811\\
\end{bmatrix}, 
\end{equation*}
\begin{equation*}
   \textbf{CB}_6^{LS} =
\begin{bmatrix}
          0  &       0   &      0   &      0\\
    0.0788   & 9.8115    &0.0528&    1.9884\\
    3.9862    &1.1829    &0.7601 &   0.0323\\
         0     &    0     &    0  &       0\\
\end{bmatrix}.
\end{equation*}

\section*{Acknowledgment}
The authors would like to thank the Associate Editor Dr. Nestor Chatzidiamantis and the anonymous reviewers
 for their constructive suggestions which have greatly
helped to improve the quality of this work. The authors are also thankful
to Mr. Vyacheslav P. Klimentyev for the publicly available codes and various discussions.


\begin{thebibliography}{00}

\bibitem{singleled_1}
S. Rajagopal, R. D. Roberts, and S. K. Lim, “IEEE 802.15.7 visible
light communication: Modulation schemes and dimming support,” \emph{IEEE
Commun. Mag.}, vol. 50, no. 3, pp. 72–82, Mar. 2012.

\bibitem{singleled_2}
M. Kavehrad, “Sustainable energy-efficient wireless application using
light,” \emph{IEEE Commun. Mag.}, vol. 48, no. 12, pp. 66–73, Dec. 2010.

\bibitem{single_led3}
A. Jovicic, J. Li, and T. Richardson, “Visible light communication: Opportunities,
challenges and the path to market,” \emph{IEEE Commun. Mag.}, vol. 51,
no. 12, pp. 26–31, Dec. 2013.


\bibitem{lisuyu6gpaper_2}
Z. Ding, X. Lei, G. K. Karagiannidis, R. Schober, J. Yuan, and V. K.
Bhargava, “A survey on non-orthogonal multiple access for 5G networks:
Research challenges and future trends,” \emph{IEEE J. Sel. Areas Commun.},
vol. 35, no. 10, pp. 2181-2195, Oct. 2017.

\bibitem{scma6g_6}
S. Chen, Y.-C. Liang, S. Sun, S. Kang, W. Cheng, and M. Peng, “Vision,
requirements, and technology trend of 6G: how to tackle the challenges
of system coverage, capacity, user data-rate and movement speed,” \emph{IEEE
Wireless Commun.}, vol. 27, no. 2, pp. 218-228, Apr. 2020.


\bibitem{starqam_5}
H. Nikopour and H. Baligh, “Sparse code multiple access,” in \emph{Proc. IEEE
24th Annu. Int. Symp. Pers.}, Indoor, Mobile Radio Commun., London,
UK, Sep. 2013, pp. 332–336.

\bibitem{starqam_18}
R. Hoshyar, F. P. Wathan, and R. Tafazolli, “Novel low-density signature
for synchronous CDMA systems over AWGN channel,” \emph{IEEE Trans.
Signal Process.}, vol. 56, no. 4, pp. 1616–1626, Apr. 2008.

\bibitem{starqam_7}
M. Taherzadeh, H. Nikopour, A. Bayesteh, and H. Baligh, “SCMA CB
design,” in \emph{Proc. IEEE 80th Veh. Technol. Conf.}, Vancouver, Canada,
Sep. 2014, pp. 1–5.


\bibitem{nomavlc_7}
S. Dimitrov, S. Sinanovic, and H. Haas, “Clipping noise in OFDM based
optical wireless communication systems,” \emph{IEEE Trans. Commun.},
vol. 60, no. 4, pp. 1072–1081, Apr. 2012.

\bibitem{nomavlc_importance}
H. Marshoud, V. M. Kapinas, G. K. Karagiannidis and S. Muhaidat, "Non-Orthogonal Multiple Access for Visible Light Communications," in \emph{IEEE Photon. Technol. Lett.}, vol. 28, no. 1, pp. 51-54, 1 Jan.1, 2016.


\bibitem{vimal_12} 
B. Lin, X. Tang, Z. Zhou, C. Lin, and Z. Ghassemlooy, “Experimental demonstration of SCMA for visible
light communications,” \emph{Opt.
Commun.}, vol. 419, pp.
36–40, July 2018.

\bibitem{vimal_13}
J. An and W.-Y. Chung, “Single-LED multichannel optical
transmission with SCMA for long range health information
monitoring,” \emph{J. Lightw. Technol.}, vol. 36, no. 23, pp. 5470–5480,
2018.

\bibitem{vimal_14}
B. Lin, X. Tang, and Z. Ghassemlooy, “A power domain sparse
code multiple access scheme for visible light communications,”
\emph{IEEE Wireless Commun. Lett.}, vol. 9, no. 1, pp. 61–64, 2019.

\bibitem{vimalpaper}
R. Mitra, S. Sharma, G. Kaddoum and V. Bhatia, "Color-domain SCMA NOMA for visible light communication", \emph{IEEE Commun. Lett.}, vol. 25, no. 1, pp. 200-204, Jan. 2021.

\bibitem{vimal_8}
S. Lou, C. Gong, Q. Gao, Z. Xu, “SCMA with low complexity symmetric CB
design for visible light communication,” in \emph{Proc. International
Conference on Communications (ICC)}, July 2018, pp. 1–6.

\bibitem{rmedpaper}
S. Hu, Q. Gao, C. Gong, Z. Xu, R. Boluda-Ruiz, and K. Qaraqe,
“Energy-efficient modulation for visible light SCMA system with signal-dependent
Noise,” in \emph{Proc. Int. Symp. Commun. Syst., Netw. Digit. Signal
Process. (CSNDSP)}, Budapest, Hungary, Jul. 2018, pp. 1–6.

\bibitem{dropaper}
Q. Gao, S. Hu, C. Gong, E. Serpedin, K. Qaraqe and Z. Xu, “Distance-range-oriented constellation design
for VLC-SCMA downlink with signal-dependent noise,” \emph{IEEE
Commun. Lett.}, vol. 23, no. 3, pp. 434–437, 2019.


\bibitem{shotnoise_ber_1}
{J. Y. Wang, J. B. Wang and Y. Wang, “Fundamental Analysis for
Visible Light Communication with Input‐Dependent Noise, Optical
Fiber and Wireless Communications”, PhD. Rastislav Róka (Ed.),
InTech, DOI: 10.5772/68019, (2017).}

\bibitem{shotnoise_ber_2}
V. Dixit and A. Kumar, ‘‘An exact BER analysis of NOMA-VLC system
with imperfect SIC and CSI,’’ \emph{AEU Int. J. Electron. Commun.}, vol. 138,
Aug. 2021, Art. no. 153864, doi: 10.1016/j.aeue.2021.153864.

\bibitem{dna_rv2}
J. M. Senior, and M. Y. Jamro, Optical Fiber Communications: Principles and Practice. 3rd ed. Prentice Hall, 2008.

\bibitem{sdnscalingfactorpaper}
S. M. Moser, “Capacity results of an optical intensity channel with
input-dependent gaussian noise,” \emph{IEEE Trans. Inform. Theory}, vol. 58,
no. 1, pp. 207-223, Jan. 2012.

\bibitem{yaseenidgn11_23}
Q. Gao, S. Hu, and Z. Xu, “Modulation designs for visible light communications
with signal-dependent noise," \emph{J. Lightwave Technol.}, vol. 34, no.
23, pp. 5516-5525, Dec. 2016.


\bibitem{yaseenidgn11}
J. Wang, H. Ge, J. Zhu, J. Wang, J. Dai, and M. Lin, “Adaptive spatial
modulation for visible light communications with an arbitrary number
of transmitters,” \emph{IEEE Access}, vol. 6, pp. 37 108–37 123, 2018.



\bibitem{shotnoise2_4}
K.-I. Ahn and J. K. Kwon, “Capacity analysis of M-PAM inverse source
coding in visible light communications,” \emph{J. Lightw. Technol.}, vol. 30,
no. 10, pp. 1399–1404, Jan. 2012.

\bibitem{shotnoise2_5}
J.-B. Wang, Q.-S. Hu, J. Wang, M. Chen, and J.-Y. Wang, “Tight bounds on channel capacity for dimmable visible
light communications,” \emph{J. Lightw. Technol.}, vol. 31, no. 23, pp. 3771–3779, Dec. 2013.


\bibitem{shotnoise2_9}
A. Chaaban, Z. Rezki, and M.-S. Alouini, “Fundamental limits of parallel optical wireless channels: capacity results and outage formulation,”
\emph{IEEE Trans. Commun.}, vol. 65, no. 1, pp. 296–311, Jan. 2017.

\bibitem{dna_los_rv2}
D. N. Anwar, R. Ahmad and A. Srivastava, "Energy-Efficient Coexistence of LiFi Users and Light Enabled IoT Devices," in \emph{IEEE Transactions on Green Communications and Networking}, doi: 10.1109/TGCN.2021.3116267.

\bibitem{shotnoise2_11}
J. Grubor, S. Randel, K.-D. D. Langer, and J. Walewski, “Broadband
information broadcasting using LED-based interior lighting,” \emph{J. Lightw.
Technol.}, vol. 26, no. 24, pp. 3883–3892, Dec. 2008.

\bibitem{shotnoise2_12}
 L. Hanzo, H. Haas, S. Imre, D. C. O’Brien,
M. Rupp, and L. Gyongyosi, ‘‘Wireless myths,
realities, and futures: From 3G/4G to
optical and quantum wireless,’’ \emph{Proc. IEEE},
vol. 100, no. Centennial Special Issue,
pp. 1853–1888, May 13, 2012.










\bibitem{vlc_idgn2}
H. Chen and Z. Xu, “A two-dimensional constellation design method for
visible light communications with signal-dependent shot noise,” \emph{IEEE
Commun. Lett.}, vol. 22, no. 9, pp. 1786–1789, 2018.


\bibitem{anandusermobility_15}
C. Bettstetter, H. Hartenstein, and X. Pérez-Costa, “Stochastic properties
of the random waypoint mobility model,” \emph{Wireless Netw.}, vol. 10,
no. 5, pp. 555–567, 2004.



\bibitem{superimposedcwmed2021_30}
Y. Lin, Y. Liu, Y. Siu, “Low complexity message passing algorithm for
SCMA system,” \emph{IEEE Communications Letters}, vol. 20, issue: 12 ,pp. 2466-
2469 Dec. 2016.

\bibitem{mpacomplexity}
A. Bayesteh, H. Nikopour, M. Taherzadeh, H. Baligh, and J. Ma, “Low
complexity techniques for SCMA detection,” in \emph{Proc. IEEE Globecom
Workshops}, San Diego, CA, USA, Dec. 2015, pp. 1-6.


\bibitem{scmapotentialchallenges_109}
W. B. Ameur, P. Mary, M. Dumay, J.-F. Hˇelard, and J. Schwoerer,
“Performance study of MPA, Log-MPA and Max-Log-MPA for an
uplink SCMA scenario,” in \emph{Proc. 26th Int. Conf. Telecommun. (ICT)},
2019, pp. 411–416.

\bibitem{factorgra_sumprodalgo}
F. Kschischang, B. Frey, and H. Loeliger, “Factor graphs and the
sum-product algorithm,” \emph{IEEE Trans. Inf. Theory}, vol. 47, no. 2, pp.
498–519, Feb. 2001.

\bibitem{singleled_23}
J. Liu, G.Wu, S. Li, and O. Tirkkonen, “On fixed-point implementation of
Log-MPA for SCMA signals,” \emph{IEEE Wireless Commun. Lett.}, vol. 5, no. 3,
pp. 324–327, Jun. 2016.


\bibitem{starqam}
L. Yu, P. Fan, D. Cai, and Z. Ma, “Design and analysis of SCMA
CB based on Star-QAM signaling constellations,” \emph{IEEE Trans. Veh.
Technol.}, vol. 67, no. 11, pp. 10543-10553, 2018.

\bibitem{klimenyvmed_genetic}
V. P. Klimentyev and A. B. Sergienko, ”SCMA CBs optimization
based on genetic algorithm,” \emph{European Wireless 2017; 23th European
Wireless Conference}, Dresden, Germany, 2017, pp. 1-6.

\bibitem{cbsurvey_papr}
M. Vameghestahbanati, I. Marsland, R. H. Gohary, and H. Yanikomeroglu,
“Multidimensional constellations for uplink SCMA systems—A comparative study,” \emph{IEEE Commun. Surveys Tuts.}, vol. 21, no. 3, pp. 2169–2194,
third quarter 2019.



\bibitem{dnaaccess_7}
J. M. Luna-Rivera, V. Guerra, J. Rufo-Torres, R. Perez-Jimenez, C. Suarez-
Rodriguez, and J. Rabadan-Borges, ``Low-complexity colour-shift keying based
visible light communications system,'' \emph{IET Optoelectron.}, vol. 9,
no. 5, pp. 191-198, Oct. 2015.

\bibitem{dnaaccess_22}
J. Nocedal and S. J. Wright, \emph{Numerical Optimization.} NewYork, NY, USA:
Springer-Verlag, 1999.

\bibitem{dnaaccess_16}
R. Chen, ``Solution of minimax problems using equivalent differentiable
functions,'' \emph{Comput. Math. Appl.}, vol. 11, no. 12, pp. 1165-1169,
Dec. 1985.

\bibitem{dnaaccess_11}
D. U. Campos-Delgado, J. M. Luna-Rivera, R. Perez-Jimenez,
C. A. Gutierrez, V. Guerra, and J. Rabadan, ``Constellation design
for color space-based modulation in visible light communications,'' \emph{Phys.
Commun.}, vol. 31, pp. 154-159, Dec. 2018.

\bibitem{dnaaccess_14}
E. Monteiro and S. Hranilovic, ``Constellation design for color-shift keying
using interior point methods,'' in \emph{Proc. IEEE Globecom Workshops},
Dec. 2012, pp. 1224-1228.

\bibitem{dnaaccess}
D. N. Anwar and A. Srivastava, “Constellation design for single photodetector based CSK with probabilistic shaping and white color balance,”
\emph{IEEE Access}, vol. 8, pp. 159 609–159 621, 2020.



\bibitem{starqam_anapaper_13}
M. K. Simon and M.-S. Alouini, \emph{Digital Communication Over Fading
Channels}, 1st ed. Hoboken, NJ, USA: Wiley, 2000.


\bibitem{scmapot_91}
Altera Innovate Asia FPGA Design Contest. (2015). \emph{5G
Algorithm Innovation Competition}. [Online]. Available: http://
www.innovateasia.com/5g/images/pdf/1st\%205G\%20Algorithm\%
20Innovation\%20Competition-ENV1.0\%20-\%20SCMA.pdf.

\bibitem{dhs_chap2_epdl}
R.O. Duda, P.E. Hart, and D.G. Stork, \emph{Pattern Classification}. John
Wiley \& Sons, 2001.

\end{thebibliography}
\end{document}